\documentclass[fleqn,usenatbib]{mnras}

\usepackage[T1]{fontenc}
\usepackage{ae,aecompl}

\usepackage{graphicx}	
\usepackage{amsmath}	
\usepackage{amssymb}	
\usepackage{caption}

\usepackage{hyperref}

\usepackage{etoolbox}
\makeatletter
\newcount\c@additionalboxlevel
\newcount\c@maxboxlevel
\patchcmd\@combinedblfloats{\box\@outputbox}{%
   \stepcounter{additionalboxlevel}%
   \box\@outputbox
}{}{\errmessage{\noexpand\@combinedblfloats could not be patched}}

\AtBeginShipout{%
   \ifnum\value{additionalboxlevel}>\value{maxboxlevel}%
     \typeout{Warning: maxboxlevel might be too small, increase to %
       \the\value{additionalboxlevel}%
     }%
   \fi
   \@whilenum\value{additionalboxlevel}<\value{maxboxlevel}\do{%
     \typeout{* Additional boxing of page `\thepage'}%
     \setbox\AtBeginShipoutBox=\hbox{\copy\AtBeginShipoutBox}%
     \stepcounter{additionalboxlevel}%
   }%
   \setcounter{additionalboxlevel}{0}%
}
\makeatother

\title[The AGN halo occupation]{Exploring the halo occupation of AGN using dark-matter cosmological simulations} 

\author[Georgakakis et al.]{
A. Georgakakis,$^{1}$\thanks{E-mail: age@noa.gr}, J. Comparat$^{2}$, A. Merloni$^{2}$, L. Ciesla$^{3}$, J. Aird$^{4}$, A. Finoguenov$^{2,5}$
\\
$^1$Institute for Astronomy \& Astrophysics, National Observatory of Athens, V.  Paulou  \& I.  Metaxa, 11532,  Greece\\ 
$^2$Max Planck Institut f\"{u}r Extraterrestrische  Physik, Giessenbachstra\ss e, 85748 Garching, Germany\\
$^3$Aix Marseille Univ, CNRS, CNES, LAM, Marseille, France\\
$^4$Dept. of Physics and Astronomy, University of Leicester, Leicester, LE1 7RH, UK\\
$^5$Department of Physics, University of Helsinki, Gustaf H\"allstr\"omin katu 2, Helsinki F-00014, Finland
}
\date{Accepted XXX. Received YYY; in original form ZZZ}

\pubyear{2018}

\begin{document}
\label{firstpage}
\pagerange{\pageref{firstpage}--\pageref{lastpage}}
\maketitle

\begin{abstract}
A semi-empirical model is presented that describes the distribution of Active Galactic Nuclei (AGN) on the cosmic web. It populates dark-matter halos in N-body simulations (MultiDark) with galaxy stellar masses using empirical relations based on abundance matching techniques, and then paints accretion events on these galaxies using state-of-the-art measurements of the AGN occupation of galaxies. The explicit assumption is that the large-scale distribution of AGN is independent of the physics of black-hole fueling. The model is shown to be consistent with current measurements of the two-point correlation function of AGN samples. It is then used to make inferences on the halo occupation of the AGN population. Mock AGN are found in halos with a broad distribution of masses with a mode of $\approx 10^{12}\,h^{-1} \, M_{\odot}$ and a tail extending to cluster-size halos. The clustering properties of the model AGN depend only weakly on accretion luminosity and redshift. The fraction of satellite AGN in the model increases steeply toward more massive halos, in contrast with some recent observational results. This discrepancy, if confirmed, could point to  a dependence of the halo occupation of AGN on the physics of black-hole fueling.      
\end{abstract}

\begin{keywords}
galaxies: active, galaxies: Seyfert, quasars: general, galaxies: haloes, X-rays: diffuse background
\end{keywords}



\section{Introduction}

In the current paradigm of structure formation the initial fluctuations in the density field of the dark matter distribution amplify with time and gravitationally collapse to form an evolving population of dark-matter halos. These are the sites where baryonic matter can condense and form light-emitting structures, such as galaxies, that can be traced by their electromagnetic radiation. In this picture, the relation  between dark and luminous matter provides insights onto the baryonic physics that are relevant to the formation of stars, the assembly of galaxies and perhaps the growth of the supermassive black-holes at their centres. 

In recent years diverse statistical methods have been developed to study the relation between dark-matter and galaxy stellar mass or luminosity. The two-point correlation function of galaxies has been extensively used in the literature to measure the halo occupation distribution of galaxies at fixed stellar mass or luminosity threshold \citep[e.g.][]{Berlind_Weinberg2002, Zehavi2005, Zheng2005, Zheng2007, Wake2011}. The weak lensing signal of background galaxies in the vicinity of well-selected samples of foreground galaxies provides a direct measure of the dark-matter halo mass distribution in bins of galaxy stellar mass or luminosity \citep[e.g.][]{Mandelbaum2006, Leauthaud2011, Leauthaud2012, Velander2014, Hudson2015}. Abundance matching methods, whereby a monotonic relation is assumed between dark-matter halo mass and galaxy stellar mass or luminosity, have also been successful in describing the link between dark and luminous matter over a wide range of masses and redshifts \citep[e.g.][]{Moster2010, Behroozi2010, Behroozi2013, Moster2017}. The results from the studies above have been discussed in the context of mechanisms for the supply of gas onto galaxies, the overall efficiency of star-formation at different environments and redshifts as well as models for quenching the star-formation in galaxies.

Feedback from Active Galactic Nuclei (AGN) is one of the processes that could modulate the star-formation in galaxies and possibly imprint its signature on the observed dark vs luminous matter relation. In that respect, the environment of AGN, i.e. the mass distribution of the dark-matter halos in which they live, may provide important clues on baryonic physics. There are indeed suggestions in the literature that the large-scale environment of active black-holes contains information on the impact of AGN winds on galaxies  \citep{Fanidakis2013}, the triggering mechanism of the observed nuclear activity \citep{Hopkins2007_LSS, Allevato2012}, and the physics of black-hole fueling \citep[e.g.][]{Fanidakis2013, Koutoulidis2013, Krumpe2015}. It is also recognized however, that important covariances exist between galaxy properties (e.g. star-formation rate, stellar mass) and position on the cosmic web \citep[e.g.][]{Coil2008, Zehavi2011, Cochrane2018}. When studying the environment of AGN the impact of their host galaxies on the observed signal needs to be accounted for in order to isolate a possible connection between dark-matter halo mass and accretion events onto supermassive black holes  \citep{Li2006, Georgakakis2014, Leauthaud2015, Mendez2016}. Controlling for these effects is not trivial. In the case of UV-bright QSOs for example, it is challenging to disentangle the stellar light of the underlying host galaxy from the AGN emission and hence, infer in an unbiased manner properties such as star-formation rate and stellar-mass \citep[e.g.][]{Ciesla2015}. These effects can be mitigated in the case of moderate luminosity and/or obscured AGN, such as those selected at X-ray wavelengths \citep{Brandt_Alexander2015}. The challenge in this case however, is the small number statistics that plague X-ray AGN samples to date and hamper detailed modeling of the mass distribution of their dark-matter halos \citep{Miyaji2011}. This issue will be addressed by the flow of new X-ray data from the eROSITA telescope \citep{Merloni2012} and longer term the {\it Athena} X-ray Observatory \citep{Nandra2013}. Although these missions will substantially improve the signal-to-noise ratio of clustering measurements, e.g. via the two-point correlation function, concerns have been raised on whether this is sufficient to provide meaningful constraints on the dark-matter halo-mass distribution of the AGN populations. Recent studies that use the largest samples of UV/optically selected QSOs to date ($\approx10^{4}-10^{5}$ objects) suggest that the standard Halo Occupation Distribution model suffers significant degeneracies for this type of sources \citep{Shen2013_cl, RodriguezTorres2017}. It is shown that very different HOD models are consistent with the two-point correlation function measurements. This suggests that independent observations must supplement the two-point correlation function statistic to constrain the halo-mass distribution of AGN \citep{Leauthaud2015} and/or new modelling approaches need to be developed to aid the interpretation of the data.

This paper presents a new semi-empirical model for the large-scale distribution of AGN, which can be used to compare against observational results, make realistic predictions for the clustering signal expected in future experiments and test observational selection effects and biases. The semi-empirical model is built upon the fundamental assumption that the clustering of AGN mirrors that of their host galaxies, i.e. there is no physical connection between accretion events and position on the cosmic web. This is motivated by recent results that highlight the importance of the host-galaxy properties of AGN for understanding their large-scale distribution \citep[e.g.][]{Leauthaud2015}. The strong assumption above is also tested in our work by comparing the predictions of our  semi-empirical model with observations of the large-scale distribution and halo occupation of X-ray selected AGN and UV-bright QSOs. The construction of our model uses the latest observational results on the galaxy occupation of accretion events \citep{Georgakakis2017_plz, Aird2018}. This step resembles studies that populate the stellar mass function of galaxies with AGN using parametric models for the specific-accretion rate distribution of active black-holes \citep{Aird2013, Caplar2015, Bernhard2018}. What differentiates our model from these works is that we couple empirical specific accretion-rate distributions with large N-body simulations  \citep{Klypin2016} that follow the assembly and evolution of dark-matter halos within a cosmological volume. This latter step allows placing AGN on the cosmic web and constructing a model of their large-scale distribution.  In that respect our methodology is similar to that developed by \cite{Conroy_White2013}  to explore the relationship between quasars, galaxies, and dark-matter halos. The utility and predictive power of such approaches can be demonstrated by generating realistic mock catalogues for upcoming observational programmes, such as the eROSITA X-ray sky (Comparat et al. in prep.), to quantify the expected level of uncertainties and test systematics in AGN clustering studies. 

Throughout we adopt a flat $\Lambda$-CDM cosmology with parameters similar to the Planck 2015 results \citep{Planck2016},  $\Omega_M=0.307$, $\Omega_{\Lambda}=0.693$,  $\rm H_0 = 67.77 \, km \, s^{-1}\, Mpc^{-1}$. The exact choice of values is imposed by the N-body simulations used in our work \citep[MulitDark][]{Klypin2016, Comparat2017}. All distance-dependent quantities are parametrized by $h=\rm H_0/100$.

\begin{figure*}
\begin{center}
\includegraphics[height=1.1\columnwidth]{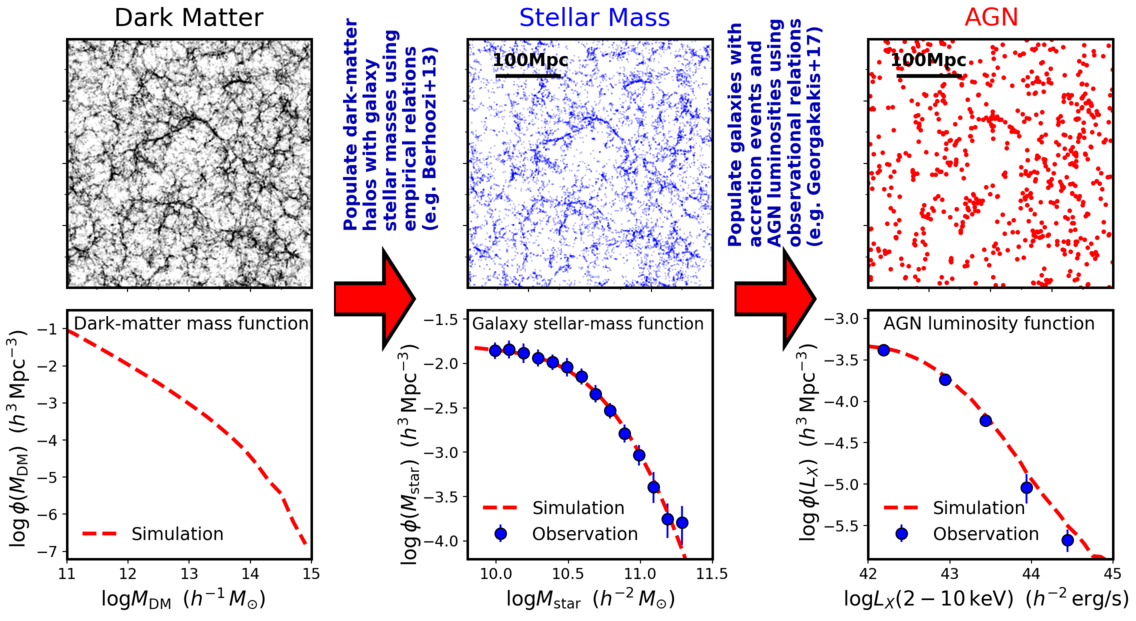}
\end{center}
\caption{Graphical representation of the semi-empirical AGN model construction. The top set of panels are slices of a cosmological simulation box from the MultiDark project \protect\citep{Klypin2016, Comparat2017} at a snapshot redshift $z=0.75$. It shows the positions of particles (e.g. left-top panel: dark matter halos; middle-top panel: galaxies; top-right panel: AGN) within the simulation box. Each particle in the simulation is represented by a dot (top-left panel: dark matter halos are shown with black; top-middle panel: galaxies are plotted in blue; top-right panel: AGN are shown in red). Darker regions mark a high density of particles, i.e. rich environments in the simulation. The construction of the AGN semi-empirical model proceeds from left to right in this graphical representation: dark-matter halos (black dots in the top-left panel) in the simulation box are populated with galaxies (blue dots in the top-middle panel) using empirical relations between dark-matter halo mass and stellar mass \protect\citep[e.g.][]{Behroozi2013}. Accretion events are then distributed in these galaxies using observationally determined probabilities that a galaxy with a given stellar mass hosts an AGN with a given accretion luminosity \citep[e.g.][]{Georgakakis2017_plz, Aird2018}. The feature of this approach is that it starts from the simulated mass function of dark-matter halos in the Universe (red-dashed curve in bottom-left panel) and reproduces by construction the observed stellar mass function of galaxies \citep[middle-bottom panel: blue circles are observations from][red-dashed curve is the model]{Moustakas2013}, and the luminosity function of AGN \protect\citep[right-bottom panel: blue points are observations from][red-dashed is the simulation]{Georgakakis2017_plz}.}\label{fig:model}
\end{figure*}

\begin{figure*}
\begin{center}
\includegraphics[height=1.2\columnwidth]{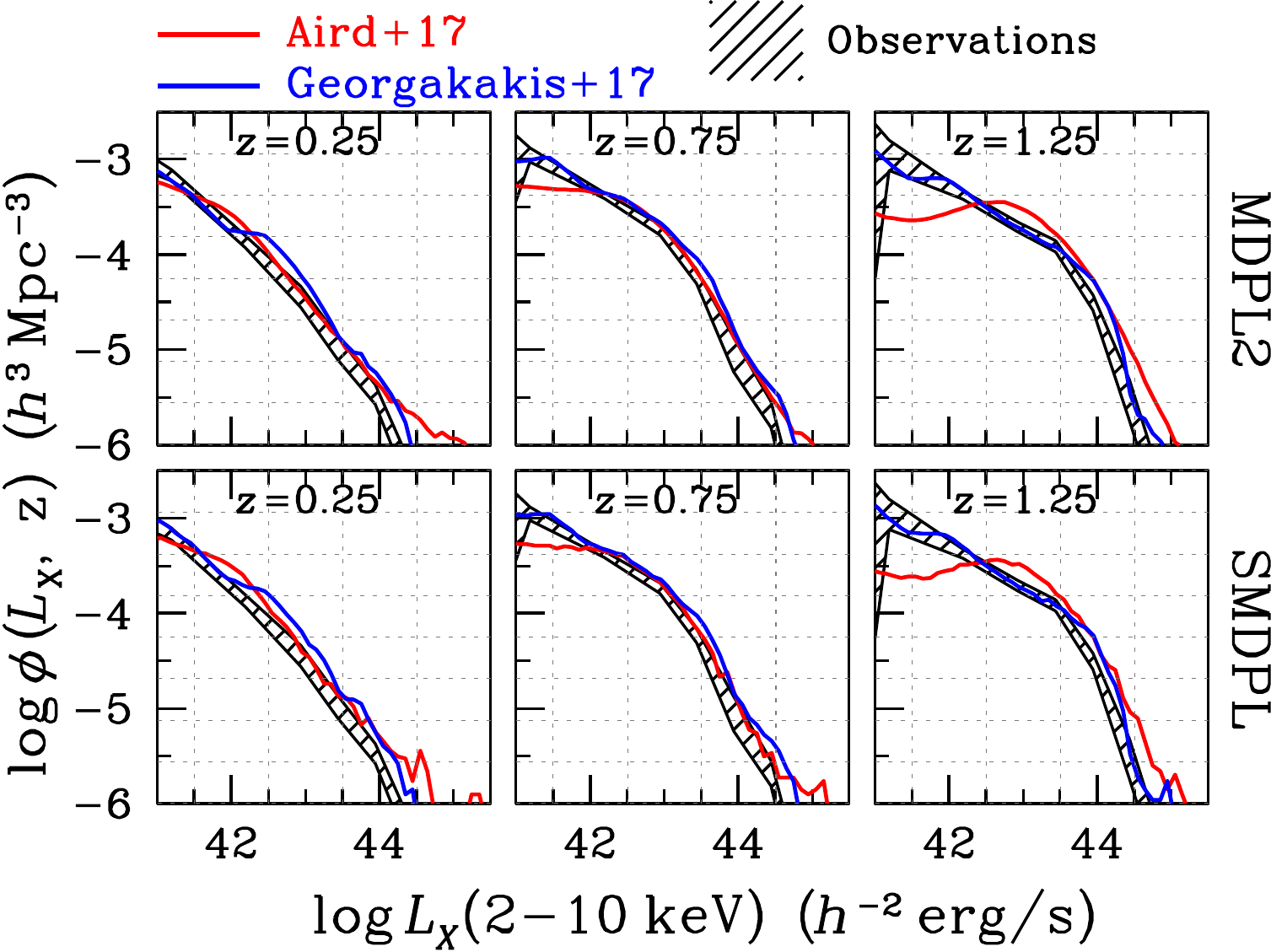}
\end{center}
\caption{Comparison of the X-ray luminosity function of AGN derived from  observational data and the semi-empirical simulations described in the text. The lower and upper set of panels correspond to the AGN mock catalogues derived  from the SMDPL and MDPL2 simulation boxes, respectively. The redshift of each simulation box is indicated on individual panels. The black-shaded regions in each panel are the observationally determined non-parametric X-ray luminosity functions of \citet[][]{Georgakakis2017_plz}. These have been derived for the redshift intervals $z=0.0-0.5$, $0.5-1.0$ and $1.0-1.5$ and are plotted at the relevant redshift panel. The  extent  of  the shaded  regions corresponds to the 90\% confidence interval of the AGN space-density, $\phi(L_X,z)$. The red and blue curves use the \protect\cite{Aird2018} and the   \protect\cite{Georgakakis2017_plz} specific accretion-rate distributions respectively, to populate mock galaxies with AGN.}\label{fig:xlf}
\end{figure*}

\section{Methodology}\label{sec:method}

The method we follow to construct mock catalogs of AGN and study their clustering properties is based on empirical relations and builds upon three recent key developments.  The proliferation of large volume and high resolution cosmological N-body simulations that describe the assembly of dark-matter halos in the Universe \citep[e.g.][]{Riebe2013}. The progress made on semi-empirical models that associate galaxies with their dark-matter halos \citep[e.g.][]{Behroozi2013}, and the state-of-the-art observational constraints on the incidence of AGN in galaxies \citep[e.g.][]{Aird2018, Georgakakis2017_plz}. The construction of the semi-empirical model for the distribution of AGN on the cosmic web is graphically demonstrated in Figure \ref{fig:model}.

Among the different models that have been developed to link galaxies with dark matter halos, the semi-empirical abundance matching approach \citep[e.g.][]{Kravtsov2004, Vale_Ostriker2004} has the advantage that with a relatively small number of parameters can populate halos with galaxies in a manner that is consistent with observational measurements of the halo vs stellar mass relation \citep[e.g.][]{Leauthaud2012, Hudson2015, Coupon2015, Ishikawa2017, Cowley2018}. The method assumes that each dark-matter halo contains a single galaxy with stellar mass (or luminosity) that is monotonically related to halo mass. N-body simulations provide information on the evolution and spatial distribution of dark-matter halos within cosmological volumes. The abundance-matching method populates these halos with stellar masses by requiring that the statistical properties of the resulting mock galaxy population (e.g. stellar mass function at given epoch, specific star-formation rates, cosmic star-formation history) match the plethora of observational data currently available \citep{Moster2010, Behroozi2010, Behroozi2013, Moster2017}.  Statistical  and  systematic  effects may also be accounted for, e.g. uncertainties in the determination of stellar masses from observations, or the scatter  between  stellar  mass  and  halo mass \citep[e.g.][]{Moster2010, Behroozi2010}. The mock galaxy catalogues produced via the abundance-matching methods above reproduce by construction the observed galaxy stellar-mass function evolution. 

Large extragalactic survey programmes \citep[e.g.][]{Brandt_Hasinger2005} that combine information from different parts of the electromagnetic spectrum have made possible the identification of large samples of AGN and the determination of key properties of their host galaxies, such as the stellar mass and the star-formation rate \citep[e.g.][]{Brandt_Alexander2015}. This data have recently been used to estimate the specific accretion-rate distribution of AGN, which measures the probability of a galaxy hosting an active nucleus with specific accretion-rate $\lambda$ \citep[e.g][]{Bongiorno2012, Bongiorno2016, Aird2012, Aird2018, Georgakakis2017_plz}. The latter quantity is defined as the ratio between the instantaneous AGN accretion-luminosity and the stellar mass of its host galaxy. Under certain assumptions, the specific accretion rate can be viewed as a proxy of the Eddington ratio of the active black hole.  The specific accretion-rate distribution of AGN provides an empirical tool to populate galaxies within a cosmological volume with specific accretion rates and hence, accretion luminosities. A feature of this approach is that the AGN luminosity function in the resulting mock catalogs matches the observed one. 

In this work we use the abundance-matching approach to populate the dark matter halos of cosmological N-body simulations with galaxies. These are then assigned accretion luminosities using observed specific-accretion rate distributions from the literature. The key assumption of the method is that there is no direct physical connection between the incidence of AGN and their position on the cosmic-web, apart from any indirect and possibly weak correlations imposed by the stellar-mass dependence of the adopted specific accretion-rate distributions (see Section \ref{sec:SARD}).

We acknowledge concerns on the ability of abundance matching methods to reconstruct accurately the observed correlation function of stellar-mass selected galaxy samples \citep{Campbell2018}. This discrepancy becomes more pronounced toward lower stellar masses ($\log [M_{\star} / h^{-2} \, M_\odot] \la 10.5$) and smaller scales ($\la \rm 1\,Mpc$). It is attributed to the low fraction of satellite galaxies produced by most abundance matching methods resulting in an underestimation of the correlation function compared to observations. Although this is an issue for the small-scale clustering of AGN predicted by the model,  the impact of this effect on our analysis is likely to be moderate. AGN are typically associated with massive galaxies close to and above the knee of the stellar mass function,  $\log [M_{\star} / h^{-2} \, M_\odot] \ga 10.5$ \citep{Georgakakis2017_plz, Aird2018}, where the discrepancies between the correlation function of abundance matching methods and observations are less pronounced or almost disappear. The preference of AGN for massive galaxies is not physical but a selection effect linked to the steepness of the specific-accretion rate distributions and the shape of the galaxy stellar mass function \citep[e.g.][]{Aird2013}.

\subsection{N-body simulations}

We use dark-matter halo simulations from the MultiDark\footnote{\url{www.cosmosim.org},  \url{www.skiesanduniverses.org}} project \citep{Riebe2013}, which currently provides the largest publicly available set of high-resolution and large-volume N-body simulations. These simulations use $3840^3$ particles in a flat $\Lambda$CDM cosmology that is consistent with Planck 2015 results \citep{Planck2016} and has cosmological parameters $\Omega_M=0.307$, $\Omega_{\Lambda}=0.693$, $\Omega_{b}=0.048$, $n_s=0.96$, $h=0.6777$ and $\sigma_{8}=0.8228$. Two independent sets of simulations with different volumes are used to explore the impact of mass-resolution effects on the results and conclusions. The Small MultiDark-Planck (SMDPL) simulations have a comoving periodic-box side of $400\,h^{-1} \,\rm Mpc$ and a mass resolution of $9.63\times10^7\,h^{-1}\,M_{\odot}$. The MultiDark-Planck 2 (MDPL2) simulations have a bigger box size, $1000\,h^{-1} \,\rm Mpc$ on the side, and a mass resolution of $1.51\times10^9\,h^{-1}\,M_{\odot}$. Details on the SMDPL and MDPL2 simulations can be found in \cite{Klypin2016} and \cite{Comparat2017}. The {\sc Rockstar} halo  finder \citep{Behroozi2013_rock} has been applied to the  SMDPL and MDPL2 simulations to identify halos and flag those (sub-halos) that lie within the virial radius of a more massive host-halo. In the rest of the paper, the mass of a dark matter halo, $M_{DM}$, is defined as the virial mass in the case of host halos and the peak progenitor virial mass for sub-halos. In the analysis that follows we use three simulation snapshots that correspond to redshifts $z=0.25$, 0.75 and 1.25. They are chosen to cover the redshift interval with the most observational measurements of the AGN clustering. 

We use the galaxy-halo model of \cite{Behroozi2013} to populate dark matter halos with stellar masses and generate mock galaxy catalogues (see Fig. \ref{fig:model}). The virial mass is used as proxy of the stellar mass in the case of host halos and the corresponding central galaxies. In sub-halos the peak progenitor virial mass is used to estimate the stellar mass of satellite galaxies. The \cite{Behroozi2013} parametric model estimates the median stellar mass at fixed dark-matter halo mass and redshift. The scatter in the stellar mass at a given dark matter halo mass is also included in our analysis. Random and systematics uncertainties that affect observationally determined galaxy stellar  masses are also parametrised and added to the model-derived stellar masses.

\subsection{Specific Accretion-Rate Distribution}\label{sec:SARD}

The specific accretion-rate distributions of AGN estimated by  \cite{Georgakakis2017_plz} and \cite{Aird2018} are independently used to assign accretion luminosities to the mock galaxies produced via the abundance-matching approach described in the previous section.  In both studies the specific accretion-rate, $\lambda$, is proportional to the quantity $L_X/M_{\star}$, i.e. the X-ray luminosity of AGN normalised to the stellar mass of their host galaxies. This quantity can be measured directly from observations and provides an estimate of how much X-rays per unit stellar mass are emitted by galaxies. The ratio $L_X/M_{\star}$ is scaled according to the relation 

\begin{equation}\label{eq:lambda}
\lambda = \frac{25\,L_X({\rm 2-10\,keV}) }{1.26\times10^{38}\;0.002 M_{\star}},
\end{equation}

\noindent where $L_X$ is units of erg/s and $M_{\star}$ is in solar units. This to make $\lambda$ resemble the AGN Eddington ratio, under the assumptions of a fixed AGN bolometric correction (the 25 factor in the equation above) and a linear scaling relation without scatter between stellar mass and black-hole mass ($M_{BH}=0.002\,M_{\star}$). Any deviations from the above assumptions modify the correspondence between the observed $L_X/M_{\star}$ ratio and the Eddington ratio and are absorbed by the overall shape of the inferred specific accretion-rate distributions.  \cite{Georgakakis2017_plz} found that the introduction of a scatter in the black-hole/stellar mass relation or of luminosity-dependent AGN bolometric corrections changes only mildly the basic characteristics of the inferred specific accretion-rate distribution. Quantities such as the AGN stellar-mass function at fixed $L_X$, which is relevant to the present study because of the $M_{\star}-M_{DM}$ relation (see previous section), remain largely unchanged. Therefore the choice of the scaling factor in Equation \ref{eq:lambda} is a second order effect to the results presented in this paper. Next we provide the most salient details of the estimation of the AGN specific accretion-rate  distributions used in our work. The reader is referred to the two relevant publications, \cite{Georgakakis2017_plz} and \cite{Aird2018},  for additional information.

\cite{Georgakakis2017_plz} presented non-parametric estimates of the specific accretion-rate distribution of X-ray selected AGN. Their starting point were deep/pencil-beam and shallow/wide-area X-ray survey data, which trace accretion events onto supermassive black holes at the centres of galaxies. These were combined with stellar masses for the host galaxies of individual X-ray sources. These are estimated using the {\sc cigale} code \citep{Boquien2018} to fit AGN templates \citep{Ciesla2015} and stellar population models to the broad-band photometry of X-ray sources and decompose the observed Spectral Energy Distributions into stellar and AGN emission. A Bayesian inference methodology was developed to constrain the non-parametric model of the specific accretion-rate distribution of AGN, by requiring that its convolution with the (fixed) galaxy stellar-mass function yields the observed number of X-ray sources in bins of luminosity, redshift and stellar mass. A key feature of the \cite{Georgakakis2017_plz} work is that systematic and random errors in e.g. photometric redshifts, X-ray luminosities and stellar masses, were accounted for in the analysis. The majority of X-ray sources used in that study were selected in the 0.5-8\,keV energy band.  In the present work we use the specific accretion-rate distributions inferred by \cite{Georgakakis2017_plz} that depend on both redshift and galaxy stellar mass. \cite{Aird2018} determined the specific accretion-rate distribution of AGN using a very different approach. Their starting point were large and deep near-infrared selected photometric galaxy catalogues, for which stellar masses and star-formation rates for individual sources were estimated using the {\sc fast} code \citep{Kriek2009} to fit AGN and galaxy templates to the observed UV-to-NIR spectral energy distributions. X-ray observations were then used to assess the mean X-ray properties in the 2-10\,keV spectral band of galaxy samples binned in redshift, stellar mass and star-formation rate. The observational constraints were fed into a flexible Bayesian mixture model to determine in a non-parametric fashion the corresponding specific accretion-rate distributions as a function of cosmic time and galaxy physical parameters (both stellar mass and star-formation rate). 

The \cite{Georgakakis2017_plz}  and \cite{Aird2018} studies overlap somewhat in the X-ray data they use, but the methods adopted to constrain the AGN specific accretion-rate distributions are quite different. They also differ in the X-ray energy band they extract information from (i.e. $0.5-8$ vs $2-10$\,keV) and the number of parameters on which the specific-accretion rate is allowed to depend on \citep[][also includes star-formation rate]{Aird2018}. Despite these differences there is good agreement between the  specific-accretion rate distributions estimated in the two independent studies \citep[see Appendix of][]{Georgakakis2017_plz}. The simulation snapshots at redshifts $z=0.25$, $0.75$ and $1.25$ are associated with the specific accretion-rate  distributions estimated by \cite{Georgakakis2017_plz} and \cite{Aird2018} in the redshift bins $z=0.0-0.5$\footnote{The lowest redshift bin of \cite{Aird2018} is $z=0.1-0.5$}, $0.5-1.0$, $1.0-1.5$, respectively. 

The assignment of accretion luminosities to mock galaxies is a stochastic process. Each mock galaxy with stellar mass $M_\star$ in a simulation box that corresponds to redshift $z$ is assigned a specific accretion-rate, $\lambda$, that is drawn randomly from the observationally determined specific accretion-rate distributions. In this process no distinction is made between central and satellite galaxies. In the case of the \cite{Aird2018} specific accretion-rate distribution we account for the dependence on the star-formation rate by splitting the mock galaxies into star-forming and quiescent in a probabilistic way by adopting the passive galaxy fraction at fixed stellar mass and redshift  parametrised by \cite{Brammer2011}. This fraction is measured for galaxies with stellar masses $M\ga 10^{10}\, M_{\odot}$. For simulated galaxies less massive than this approximate limit we extrapolate the analytic relation of \cite{Brammer2011}. The specific accretion rate associated with a galaxy is then converted to accretion luminosity at X-ray wavelengths via Equation \ref{eq:lambda}, i.e. this conversion is internally consistent with the definition of $\lambda$ in \cite{Georgakakis2017_plz} and \cite{Aird2018}. Because of the shape of the specific accretion-rate distributions, which decrease rapidly with increasing $\lambda$, the majority of mock galaxies are assigned low specific-accretion rates, which in turn translate typically to AGN luminosities $L_X\rm <10^{39}\, erg \, s^{-1}$. Both the \cite{Georgakakis2017_plz} and the \cite{Aird2018} studies use the X-ray emission as AGN diagnostic and therefore their specific accretion-rate distributions reproduce the AGN X-ray luminosity function at different redshifts. This is demonstrated in Figure \ref{fig:xlf}, which compares the X-ray luminosity function inferred from the AGN mocks with observational measurements. The level of agreement between the observed and the simulated luminosity functions in Figure \ref{fig:xlf} is limited by differences in the galaxy stellar-mass function adopted by the abundance-matching model of \cite{Behroozi2013} and the observational studies of \cite{Georgakakis2017_plz} and  \cite{Aird2018}. The flattening at $L_X \rm \la 10^{42} \, erg \, s^{-1}$ of the reconstructed X-ray luminosity function using the \cite{Aird2018} specific accretion-rate distributions is related to the way these authors correct for the X-ray emission associated with the formation of new stars in galaxies rather than accretion events onto a supermassive black-hole \citep{Aird2017}. In any case the position of the turnover is close to the luminosity cut typically adopted by observers to select clean AGN samples and avoid contamination from star-forming galaxies. The different energy bands used by \citet[][mostly 0.5-8\,keV]{Georgakakis2017_plz} and  \citet[][2-10\,keV]{Aird2018} are also responsible to some level for the variations in the reconstructed  X-ray luminosity functions derived by using the specific accretion-rate distributions from these works. 

\begin{figure}
\begin{center}
\includegraphics[height=0.9\columnwidth]{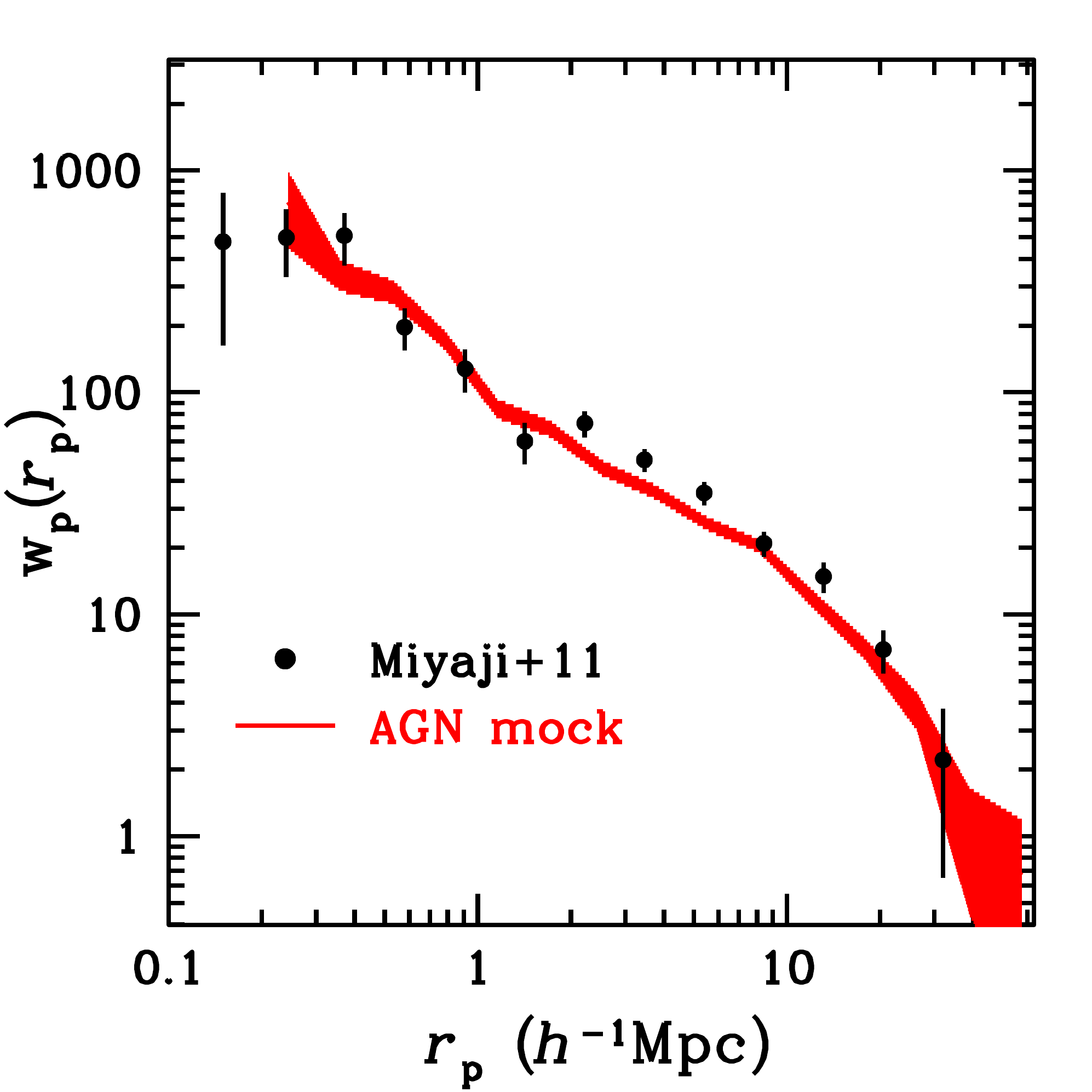}
\end{center}
\caption{The projected cross-correlation function, $w_p$, of RASS AGN and LRGs is plotted as a function of scale, $r_p$. The data-points are the observational results of \protect\cite{Miyaji2011} for their full sample of RASS AGN with $\log L_X(\rm 0.1-2.4\,keV)>43.7$ (units $\rm erg\,s^{-1}$).  The red curve corresponds to the projected cross-correlation functions for the mock RASS AGN and LRGs in the MDPL2 simulation box (see Appendix \ref{appendix:lc_miyaji}). The width of the shaded regions correspond to the $1\,\sigma$ uncertainties determined using jackknife resampling. The mock AGN catalogue is constructed using the \protect\cite{Georgakakis2017_plz} specific accretion rate distributions. 
}\label{fig:wp_agnlrg}
\end{figure}

\begin{figure}
\begin{center}
\includegraphics[height=0.9\columnwidth]{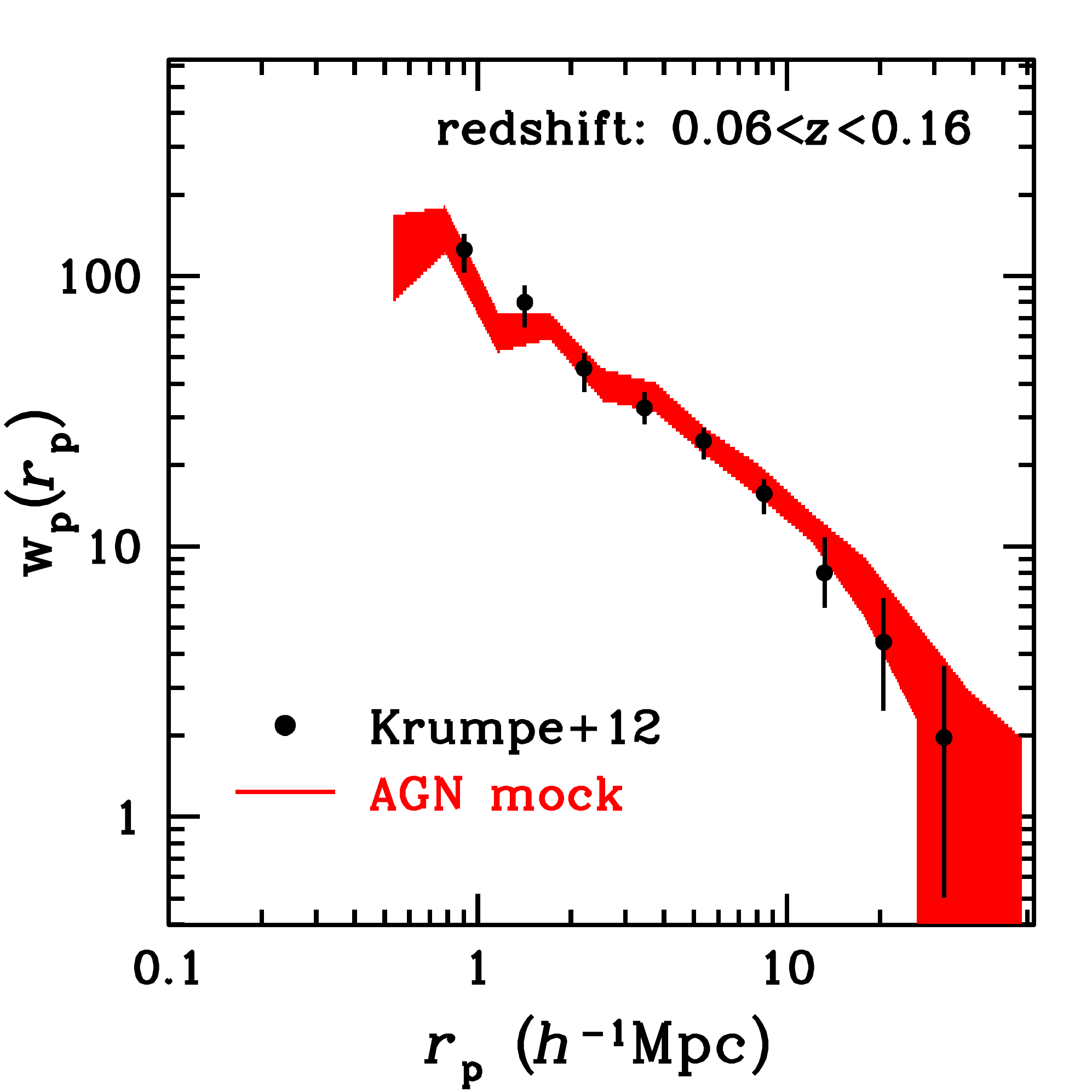}
\end{center}
\caption{The projected cross-correlation function of the RASS AGN and the SDSS Main galaxy sample selected in the redshift interval $0.06<z<0.16$ and absolute magnitude range $-20.0 < M_r <21.0$\,mag. The data-points are the observational results of \protect\cite{Krumpe2012}.  The red curve corresponds to the simulated data described in the Appendix \ref{appendix:lc_sdssmain}. The width of the shaded regions correspond to the $1\,\sigma$ uncertainties determined using jackknife resampling. The mock AGN catalogue is constructed using the \protect\cite{Georgakakis2017_plz} specific accretion-rate distributions. 
}\label{fig:wp_rasssdss}
\end{figure}

\begin{figure}
\begin{center}
\includegraphics[height=0.9\columnwidth]{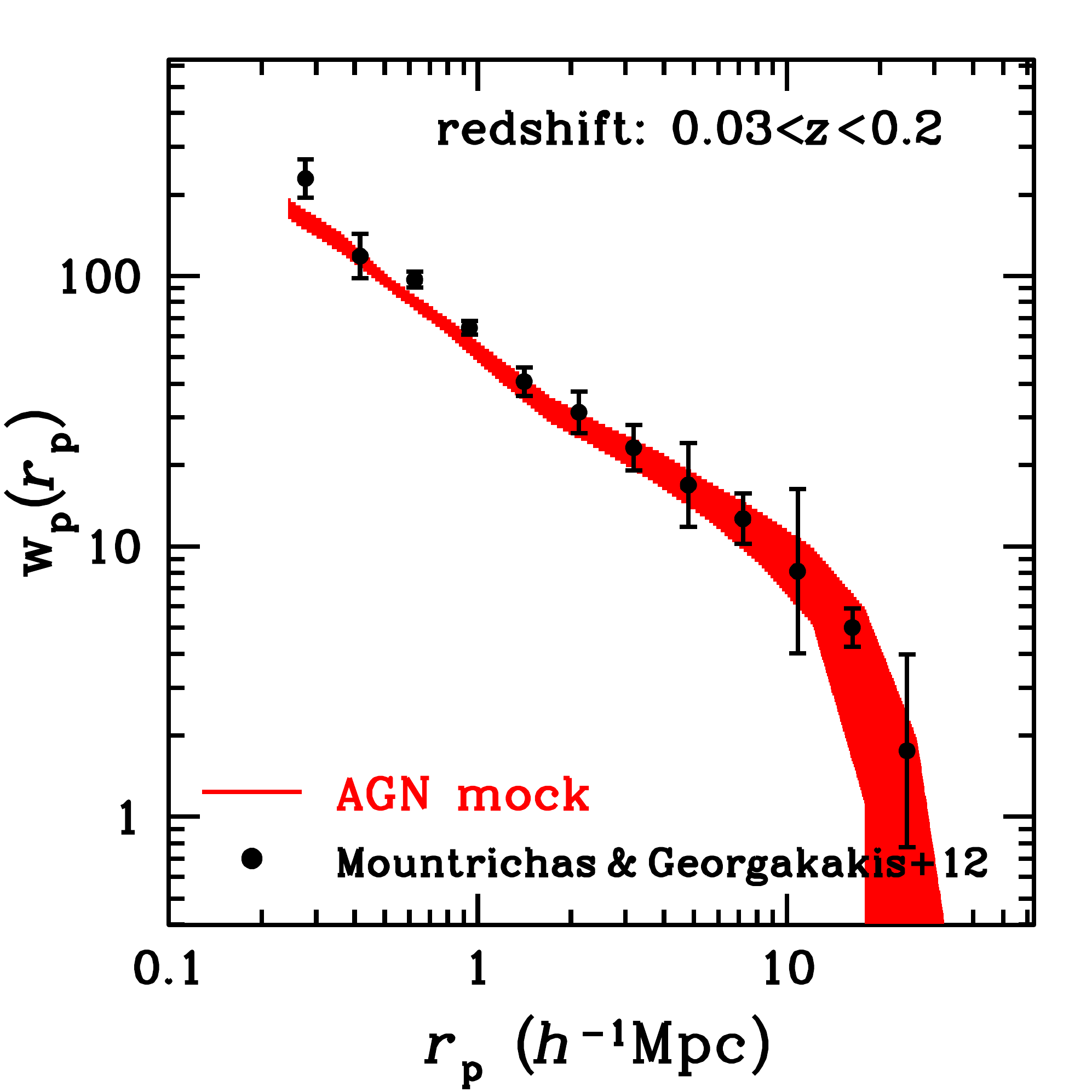}
\end{center}
\caption{The projected cross-correlation function of the SDSS Main galaxy sample and the AGN selected in the 0.5-10\,keV band of the XMM/SDSS survey \protect\citep{Georgakakis_Nandra2011}. The redshift interval of the two samples is $0.02<z<0.2$. The data-points are the observational results presented by \protect\cite{Mountrichas2012}.  The red curve is the cross-correlation function of the simulated XMM/SDSS AGN and SDSS Main galaxy sample described in Appendix \ref{appendix:lc_sdssmain}. The width of the shaded regions correspond to the $1\,\sigma$ uncertainties determined using jackknife resampling. The mock AGN catalogue is constructed using the \protect\cite{Georgakakis2017_plz} specific accretion-rate distributions. 
}\label{fig:wp_xmmsdss}
\end{figure}

\begin{figure}
\begin{center}
\includegraphics[height=0.9\columnwidth]{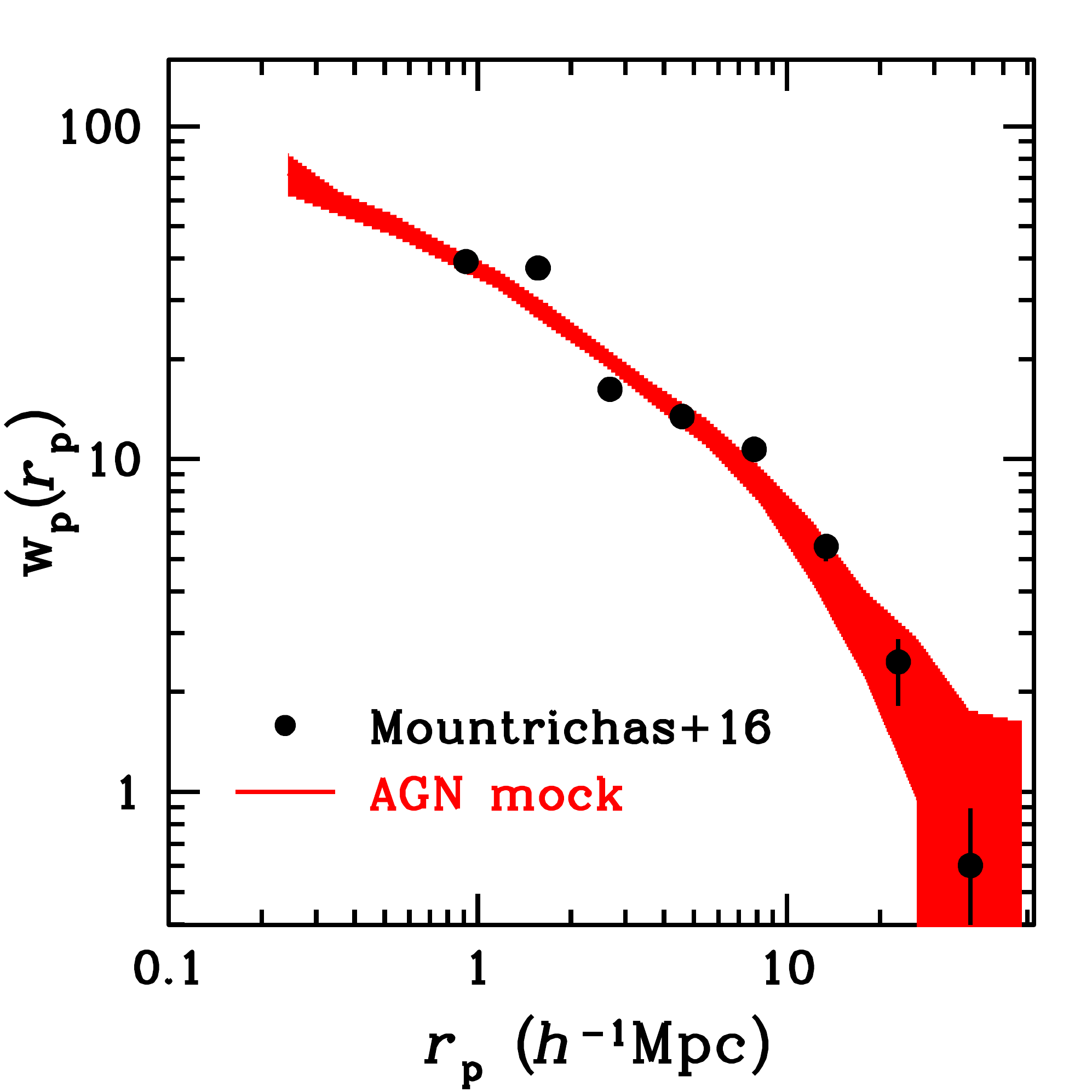}
\end{center}
\caption{The projected cross-correlation function, $w_p$, of the XMM-XXL AGN and the VIPERS galaxies is plotted as a function of scale, $r_p$. The data-points are the observational results of \protect\cite{Mountrichas2016}.  The red curve corresponds to the mock XMM-XXL AGN and VIPERS galaxies in the MDPL2 simulation box (see Appenix \ref{appendix:lc_vipers}). The width of the shaded regions correspond to the $1\,\sigma$ uncertainties determined using jackknife resampling. The mock AGN catalogue is constructed using the \protect\cite{Georgakakis2017_plz} specific accretion-rate distributions. 
}\label{fig:wp_xxlvipers}
\end{figure}

\begin{figure}
\begin{center}
\includegraphics[height=0.9\columnwidth]{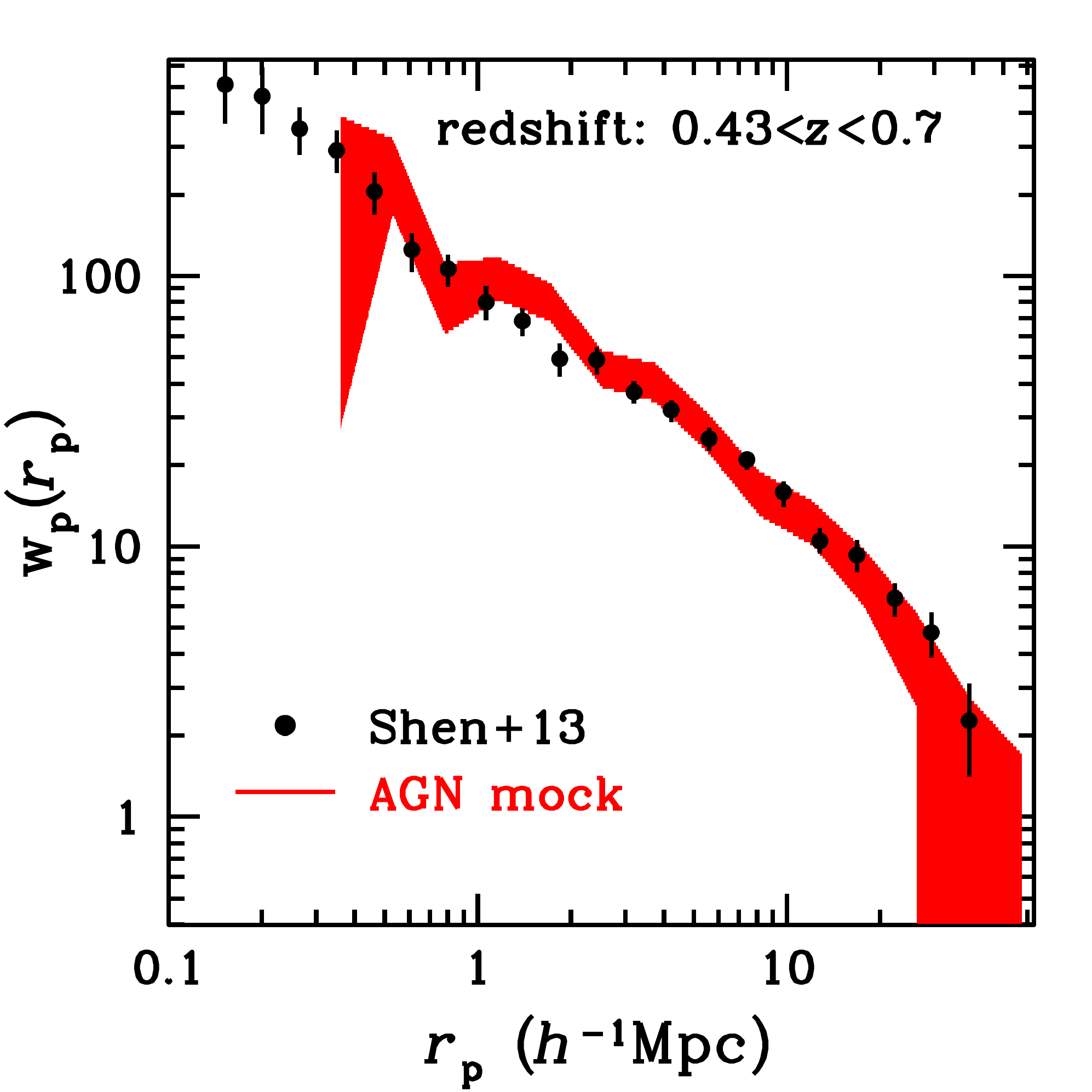}
\end{center}
\caption{The projected cross-correlation function, $w_p$, of the SDSS-DR7 QSOs and BOSS/CMASS galaxies is plotted as a function of scale, $r_p$. The data-points are the observational results of \protect\cite{Shen2013} for their full QSO sample.  The red shaded region corresponds to the mock QSO/CMASS-galaxy  cross-correlation function (see Appendix \ref{appendix:lc_shen} for details). The width of the shaded regions correspond to the $1\,\sigma$ uncertainties determined using jackknife resampling. At small scales ($r_p\la 0.5\,h^{-1}\rm \, Mpc$) there are not enough pairs in the mock catalogue to estimate the cross-correlation function. This is related to the fact that the mock light-cone is smaller in area ($\rm 706\,deg^2$) compared to the real observations of \protect\citet[][$\rm 6248\,deg^2$]{Shen2013}. The reader is referred to the Appendix \ref{appendix:lc_shen} for details. The mock AGN catalogue is constructed using the \protect\cite{Georgakakis2017_plz} specific accretion-rate distributions. 
}\label{fig:wp_shen}
\end{figure}

\section{Results}

\subsection{The 2-point correlation function of mock AGN}

We first explore how the large-scale distribution of simulated AGN compares with observational results. A forward-modeling approach is adopted for this exercise. Mock observations that mimic real data are generated by the model and are then used to estimate the same quantities that observers measure to quantify the clustering of AGN. The comparison is with observational studies that select AGN at both X-ray and UV/optical wavelengths, with emphasis on the former. This is because observations at high energies sample active supermassive black holes in galaxies with a wide range of accretion luminosities and redshifts \citep[e.g.][]{Brandt_Alexander2015}, and hence provide information on the luminosity dependence of the AGN clustering over a range of cosmic times. Also, the selection function of X-ray surveys is relatively easy to quantify and reproduce in simulations to control against potential sample-selection biases. The statistic we use as diagnostic of the AGN clustering is the 2-point correlation function that has been extensively used in the observational literature. 

Many observational studies choose to infer the clustering properties of AGN by estimating the 2-point cross-correlation function with a tracer population of galaxies \citep[e.g.][]{Coil2009, Krumpe2012, Mountrichas2013}. The motivation for this choice is practical. Random and cosmic-variance errors are minimised when estimating the cross-correlation of a typically sparse and small X-ray AGN sample with a larger tracer-sample of galaxies \citep[e.g.][]{Coil2007}. The calculation of the AGN/galaxy cross-correlation function in simulations requires knowledge on the halo distribution of both the AGN and the galaxies. For the latter population this is possible if there is observational constraints on its Halo Occupation Distribution (HOD), which can then be applied to the dark matter halos in the simulated box to create tracer galaxy mocks. 

We compare the simulation results with observational studies on the AGN/galaxy cross-correlation function, for which information on the HOD of the  galaxy tracer-population is available. \cite{Miyaji2011} estimate the cross-correlation function between the AGN in the RASS \citep[ROSAT All Sky Survey][]{Voges1999} and the SDSS Luminous Red Galaxies \citep[LRGs][]{Eisenstein2001} with $g$-band absolute magnitude $M_g<-21.2$\,mag in the redshift interval $z=0.16-0.36$. \cite{Krumpe2012} built on the work of \cite{Miyaji2011} to estimate the cross-correlation function between RASS AGN and galaxies in the redshift interval $0.06<z<0.5$. Here we focus on the \cite{Krumpe2012} results that use the Main Galaxy Sample of the Sloan survey \citep{Strauss2002} at redshifts $0.06<z<0.16$. \cite{Mountrichas2012} also use the Main Galaxy Sample of the SDSS to study the clustering of low and moderate luminosity AGN ($L_X\approx10^{42}\rm \, erg \, s^{-1}$) in the serendipitous XMM/SDSS survey \citep{Georgakakis_Nandra2011}. \cite{Mountrichas2016} cross-correlated AGN selected in the equatorial field of the shallow XMM-XXL survey \citep[]{Pierre2016, Liu2016}  with galaxies from the VIPERS  \citep[Vimos Public  Extragalactic  Survey,][]{Guzzo2014} sample in the redshift interval $z=0.5-1.2$. In addition to the X-ray AGN samples above we also compare the mocks with the observed clustering properties of the UV-bright QSO sample presented by \cite{Shen2013}. They measured the cross-correlation function between SDSS-DR7 QSOs \citep{Schneider2010} and the SDSS-DR10 CMASS galaxies \citep[i.e. "constant mass",][]{Dawson2013} in the redshift interval $z=0.3-0.9$. 

The reproduction in the simulations of the selection function of the AGN and galaxy samples above requires redshift information for individual mock sources, i.e. distances from a fiducial observer. For that purpose the simulation boxes need to be projected to the sky to produce light-cones \citep[e.g.][]{Fosalba2008}, which can then be treated as mock observations of the Universe. Appendices A to D describe the construction of the light cones from the simulation boxes for the AGN and galaxy samples described above, RASS-AGN and SDSS-LRGs or SDSS Main Galaxies, XMM-XXL AGN and VIPERS galaxies, XMM/SDSS AGN and SDSS Main Galaxies, SDSS-DR7 QSOs and CMASS galaxies. There are discrepancies between the redshift distribution of mock and observed AGN for some of the samples above (XMM/SDSS, Krumpe et al. 2012 sample), which indicate residual selection effects that are not accounted for by our methodology (e.g. X-ray spectral shape, spectroscopic follow-up selections etc). The impact of these discrepancies on the 2-point correlation function is investigated Appendix \ref{appendix:dndz} and is found to be small. This is because of the relatively narrow redshift range of the samples used in our analysis. Differences in the distribution of redshifts between observations and simulations within these intervals are a second order effect in the $w_p$ calculation.
The light-cones are used to estimate the projected AGN/galaxy cross-correlation functions in redshift-space. The uncertainties are calculated using the jackknife resampling technique. The simulated light-cone is first split into $N_{JK}$ equal-area subregions (typically 30-100). The correlation function is then estimated $N_{JK}$ times from the $N_{JK} - 1$ subregions, i.e. by excluding one subregion at a time. The  $N_{JK}$ correlation functions are then used to determine the corresponding co-variance matrix \citep[e.g.][]{Krumpe2010}. The uncertainties of the projected correlation function at a given scale are the diagonal elements of this matrix. 

Figures \ref{fig:wp_agnlrg},  \ref{fig:wp_rasssdss}, \ref{fig:wp_xmmsdss},  \ref{fig:wp_xxlvipers}  and \ref{fig:wp_shen} compare the projected correlation function estimated from the light-cones described in the Appendix with the corresponding observational results. In these figures the AGN mock catalogues are constructed using the \cite{Georgakakis2017_plz} specific accretion-rate distributions. The \cite{Aird2018} $\lambda$-distributions  produce very similar results and are not shown for the sake of brevity and clarity. The agreement between model and observations in Figure \ref{fig:wp_agnlrg}--\ref{fig:wp_shen}  is remarkable and shows that the mock catalogues of Section \ref{sec:method} are consistent with at least a subset of the current observational constraints on the large-scale distribution of X-ray selected AGN and UV-bright QSOs at $z<1$. Therefore the scheme of populating dark-matter halos with galaxies and then assigning them accretion luminosities based on empirical relations generates AGN populations with realistic clustering properties, as measured by the two-point correlation-function statistic. Next we use the semi-empirical model to explore the halo occupation properties of mock-AGN and make inferences about the real Universe.

\begin{figure*}
\begin{center}
\includegraphics[width=2\columnwidth]{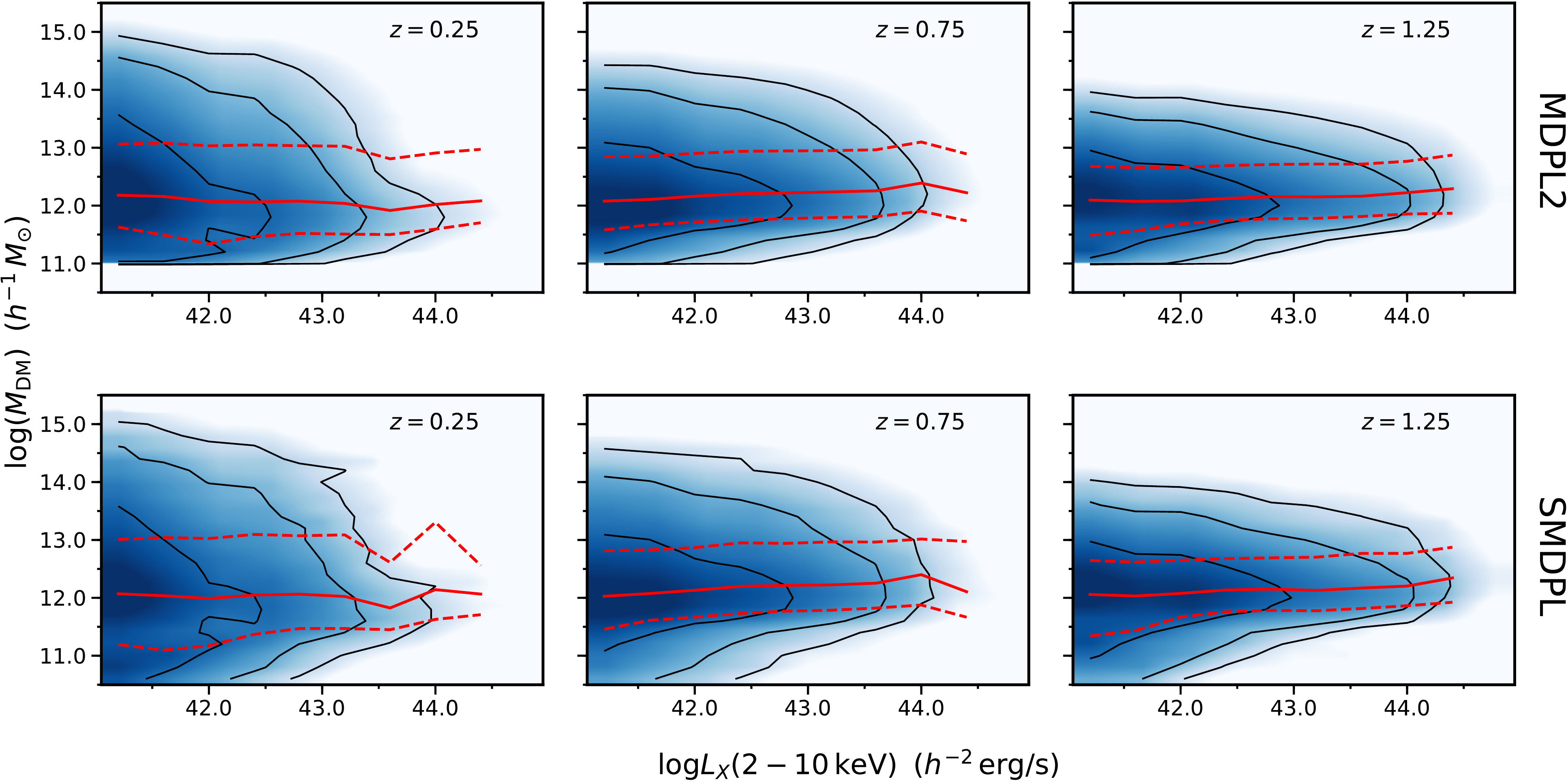}
\end{center}
\caption{Mass distribution of the dark-matter halos that host AGN as a function of X-ray luminosity. The top set of panels is for the MDPL2 simulations boxes ($1\,h^{-1}\rm \,Mpc$) and the lower set of panels corresponds to the smaller ($0.4\,h^{-1}\, \rm Mpc$ box size) SMDPL simulation. The panels in each row correspond to the redshifts of the simulation boxes used in this work, $z=0.25$, 0.75 and 1.25. The contour levels are chosen to enclose 68, 95 and 99\% of the total number of mock AGN in the simulation box. The red solid line is the median of the distribution at fixed X-ray luminosity. The dotted lines mark the 1\,sigma scatter (16th and 84th percentiles) around the median. The results are for the \protect\cite{Georgakakis2017_plz} specific accretion-rate distribution. 
}\label{fig:dmhlx2d}
\end{figure*}

\begin{figure*}
\begin{center}
\includegraphics[width=2\columnwidth]{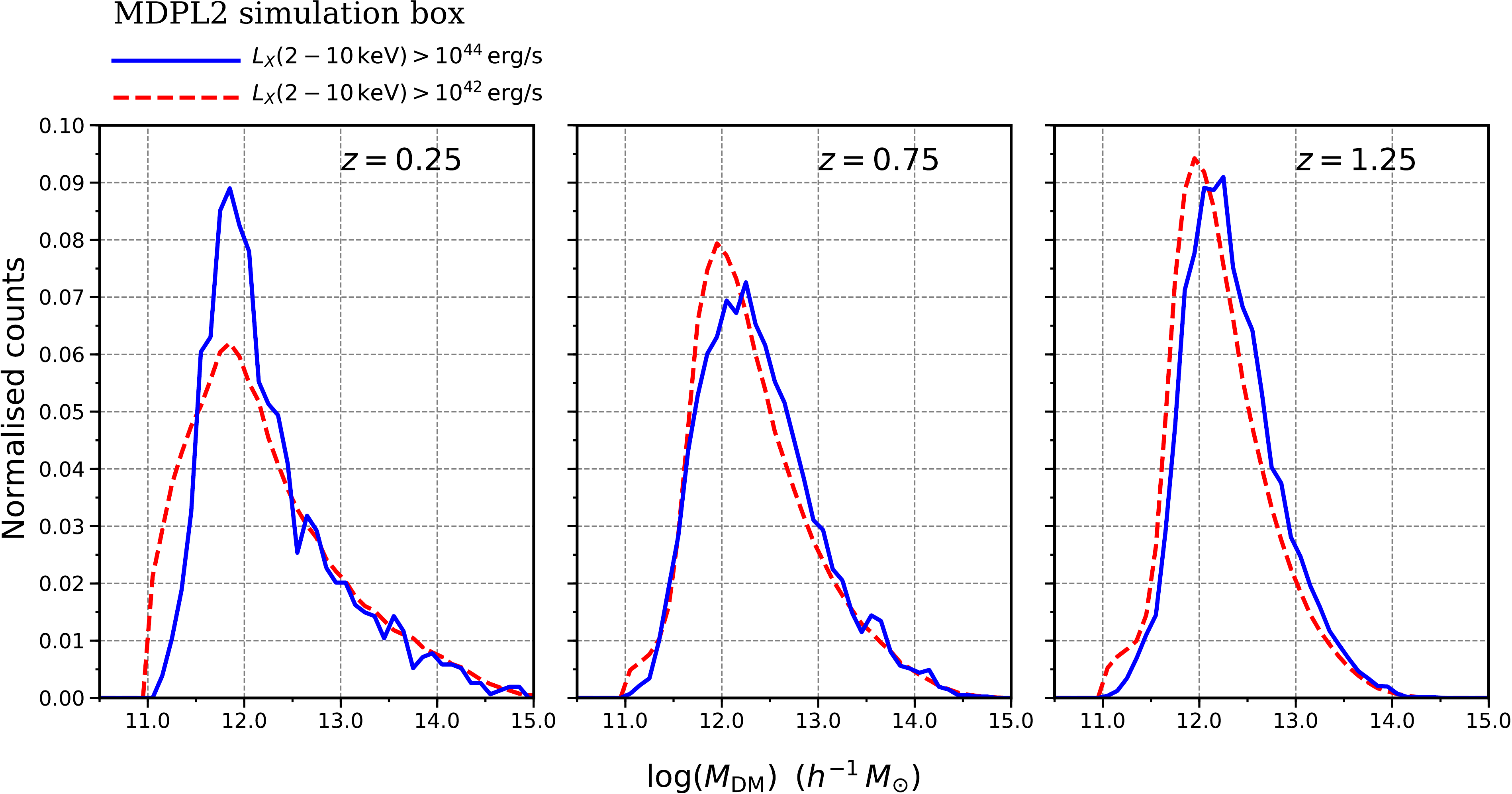}
\end{center}
\caption{The mass distribution of the dark-matter halos in the MDPL2 simulation that host AGN with X-ray luminosities $L_X(\rm 2-10\,keV)>10^{42}\, \rm erg\,s^{-1}$ (red-dashed line) and $>10^{44}\,\rm erg\,s^{-1}$ (blue-soled line). Each panel corresponds to the redshift of the simulation boxes used in this work, $z=0.25$, 0.75 and 1.25. For clarity we only show results for the MDPL2 simulation box using the \protect\cite{Georgakakis2017_plz} specific accretion-rate distributions. Using the smaller SMDPL-simulation box or the \protect\cite{Aird2018} specific accretion-rate distributions to populate halos with AGN does not change the shape of the plotted histograms.}\label{fig:dmhlx1d}
\end{figure*}

\begin{figure*}
\begin{center}
\includegraphics[width=2\columnwidth]{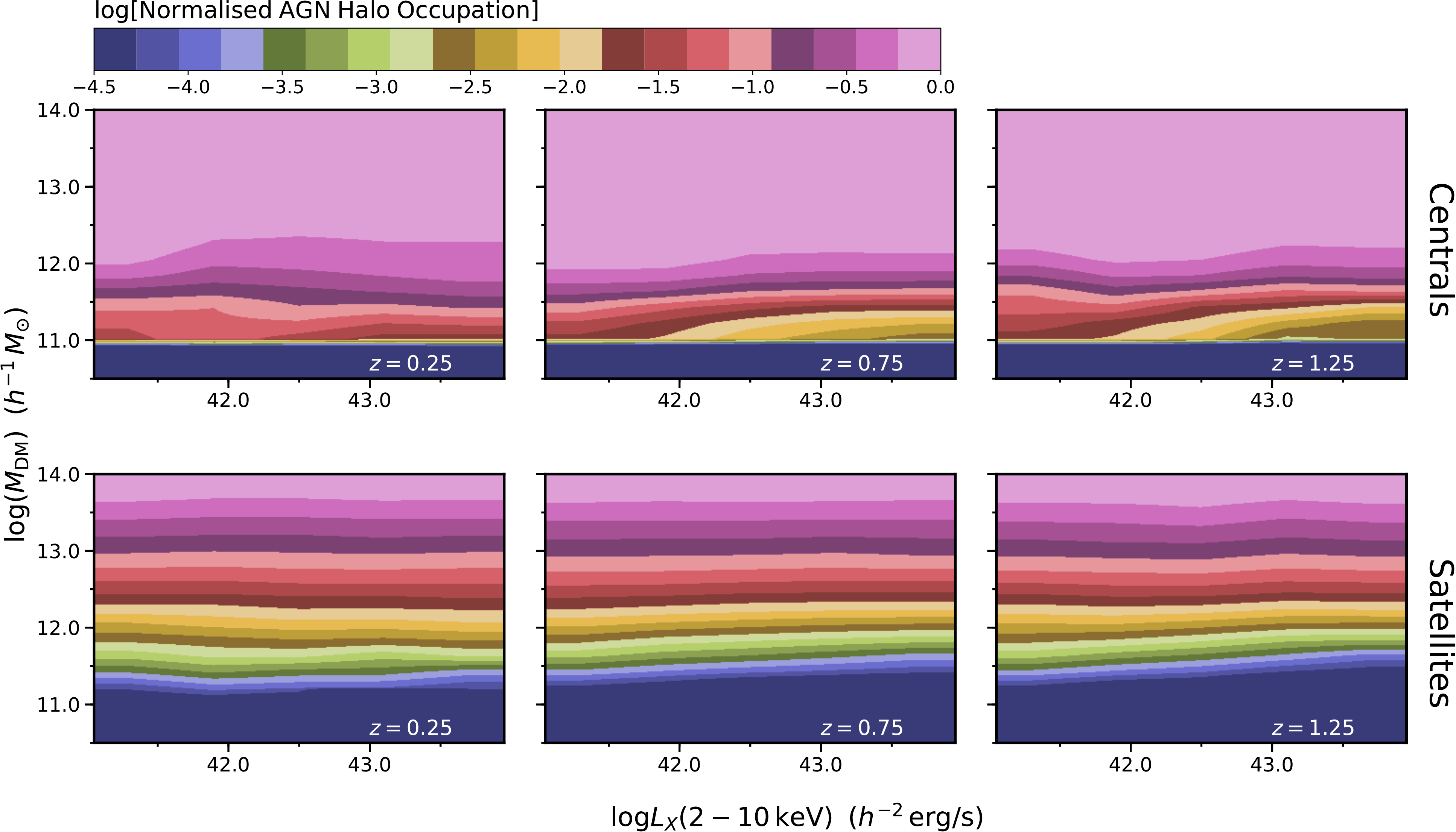}
\end{center}
\caption{The distribution of the mock-AGN halo occupation in the 2-dimensional space of halo mass and X-ray luminosity. AGN associated with central and satellites galaxies are plotted separately on the top and bottom set of panels respectively. The panels in each row correspond to the redshifts of the simulation boxes used in this work, $z=0.25$, 0.75 and 1.25. The different colours correspond to different values (in log) of the halo occupation as indicated by the colourbar at top. At fixed $L_X$ the halo occupation of both centrals and satellites  is normalized to a maximum value of unity to facilitate the visualisation of the results. The halo occupation is estimated for AGN in the  MDPL2 simulation boxes based on the \protect\cite{Georgakakis2017_plz} specific accretion-rate distributions.
}\label{fig:HOD2d}
\end{figure*}

\subsection{The halo occupation distribution of mock AGN}

Figure  \ref{fig:dmhlx2d} plots the distribution of mock AGN in the 2-dimensional space of halo mass and X-ray luminosity at different redshift intervals. It shows a broad range of dark-matter halos for the AGN at a given luminosity cut. The median of the distribution is about  $10^{12}\,h^{-1}\,M_{\odot}$ and the scatter at fixed luminosity and redshift is $\approx0.5-1$\,dex in mass. For visualisation purposes we also slice the 2-dimensional space of Figure \ref{fig:dmhlx2d} to generate histograms of the AGN halo-mass at fixed accretion luminosity cuts. These are plotted in Figure \ref{fig:dmhlx1d}. They further demonstrate that the most common environment of mock AGN are halos with masses of about $10^{12}\,h^{-1}\,M_{\odot}$ and that there is a strong tail in the distributions that extends to cluster-scale halo masses, $10^{14}-10^{15}\,h^{-1}\,M_{\odot}$. These generic trends are independent of AGN accretion-luminosity or redshift. They are also insensitive to the resolution of the dark-matter simulations. Figure \ref{fig:dmhlx2d}  also shows that the simulation resolution becomes important for detailed studies of low-luminosity AGN  $\log [L_X/(\rm erg\,s^{-1})] < 42.5$. At these luminositites the 95th percentile contour of the AGN population in the SMDPL simulations extends to dark-matter halo masses below about $10^{11}\,h^{-1}\,M_{\odot}$, i.e. the approximate resolution limit of the MDPL2 simulations adopted in this work. A small level of incompleteness is therefore expected in the larger MDPL2 simulations in the case of low-luminosity AGN.

\begin{figure}
\begin{center}
\includegraphics[height=0.8\columnwidth]{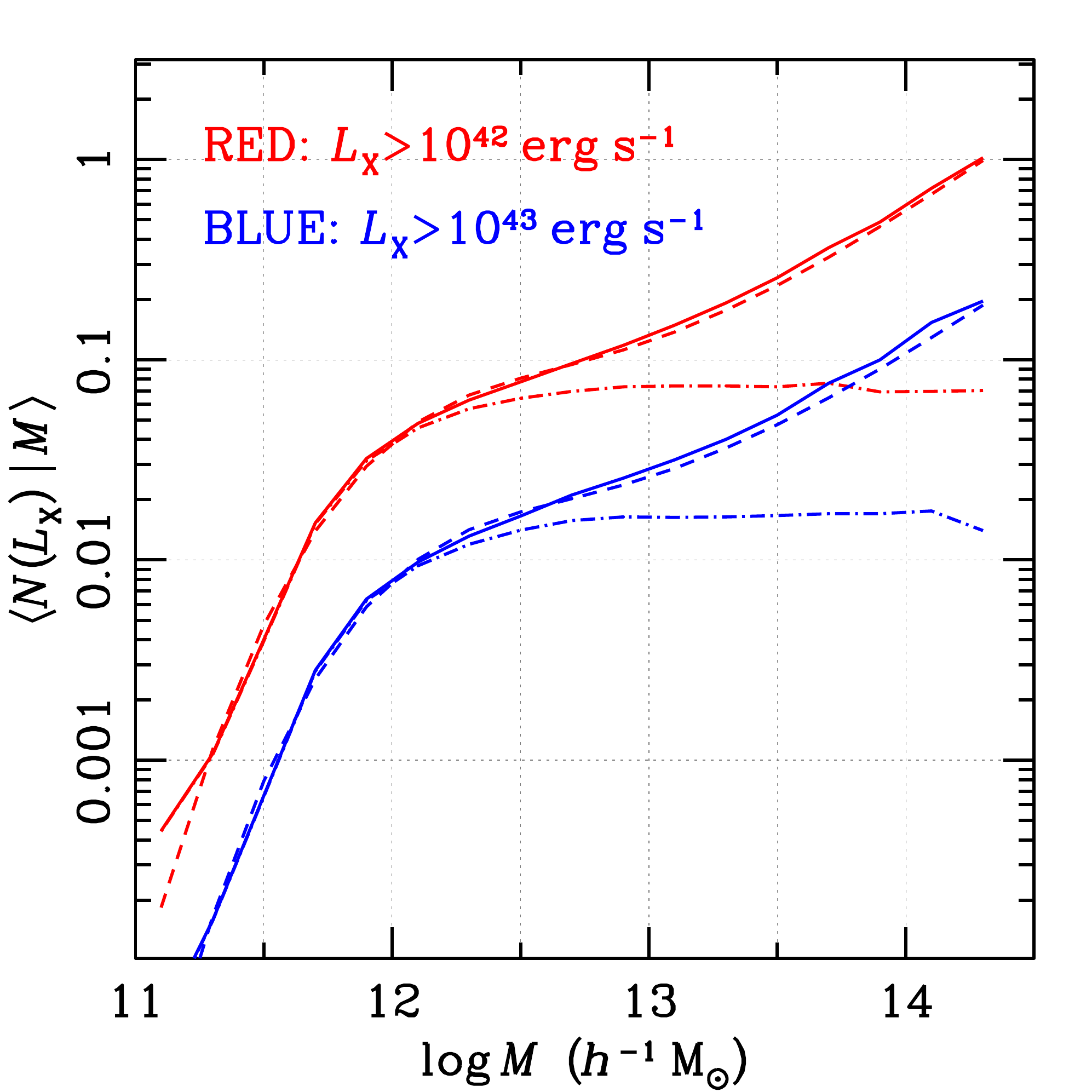}
\end{center}
\caption{Examples of the halo occupation of mock AGN (using the \protect\cite{Georgakakis2017_plz} specific accretion-rate distributions) drawn from the MDPL2 simulation box at $z=0.75$. The red and blue curves correspond to AGN with luminosities brighter than $L_X(\rm 2-10\,keV)>10^{42} \,erg\,s^{-1}$ and  $>10^{43}\rm  \,erg\,s^{-1}$, respectively. The solid line is the total halo occupation (central and satellites) measured from the simulations. The dash-dotted line corresponds to the central AGN occupation only. The dashed curves show the corresponding best-fit HOD parametric model described in the text.
}\label{fig:hod_fit}
\end{figure}

\begin{figure}
\begin{center}
\includegraphics[height=1.3\columnwidth]{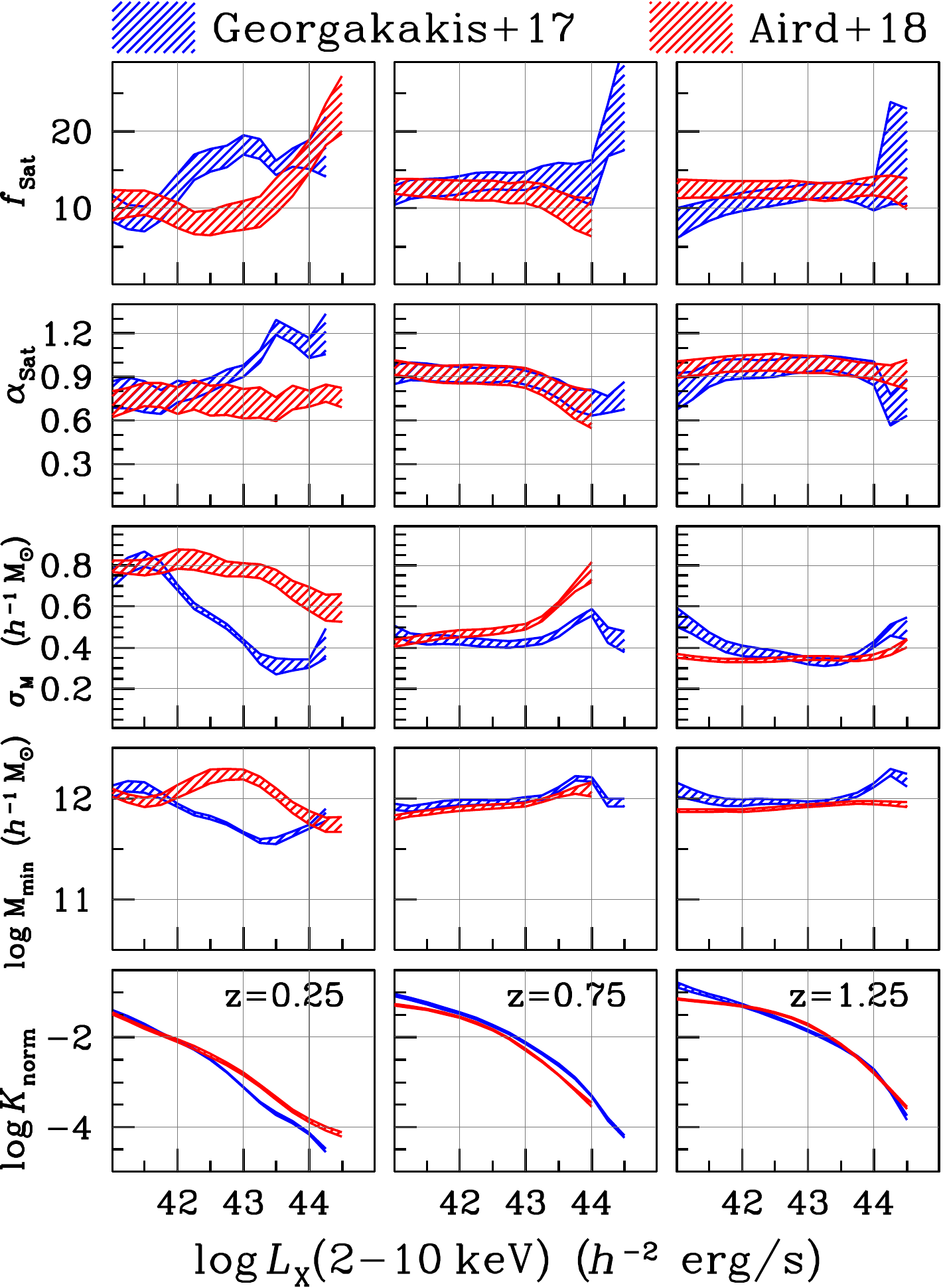}
\end{center}
\caption{Best-fit HOD parameters as function of X-ray luminosity and redshift for the mock AGN in the MDPL2 simulation boxes. Each panel row corresponds to a different parameter of the model that describes the AGN halo occupation. Each panel-column corresponds to the redshift of the MDPL2 simulation boxes, $z=0.25$, 0.75 and 1.25. The blue and red shaded regions correspond to the halo occupation parameters derived from the AGN mocks constructed using the \protect\cite{Georgakakis2017_plz} and \protect\cite{Aird2018} specific accretion-rate distributions.}\label{fig:hod_params}
\end{figure}

The distributions plotted in Figures \ref{fig:dmhlx2d} and \ref{fig:dmhlx1d} are modulated by the overall shape of the dark-matter halo-mass function. Dividing this out yields the halo occupation of AGN, i.e. the probability of an accretion event in a halo of a given mass. The distribution of this quantity in the 2-dimensional space of halo mass and accretion luminosity is shown in Figure \ref{fig:HOD2d}. The AGN occupation at a given $L_X$-bin in this figure is normalized to a maximum value of unity to make the halo-mass dependence of this quantity clearer. AGN associated with the satellite and central galaxies of a halo are plotted separately.  For the former population the probability of an accretion event at fixed accretion luminosity is a monotonic function of parent-halo mass, increasing toward higher masses. The probability of a central AGN at fixed $L_X$-limit also increases with halo mass but then saturates and remains constant above a mass limit that roughly corresponds to $10^{12}\,h^{-1}\,M_{\odot}$. These trends are reminiscent of the halo occupation of galaxies \citep[e.g.][]{Berlind_Weinberg2002, Zheng2005}.  We therefore choose to parametrise the AGN occupation using functional forms similar to those adopted in galaxy studies. We define the AGN occupation, $\langle N(L_X)\;|\;M\rangle$, as the mean number of AGN brighter than $L_X$ in halos of mass $M$. We assume that this quantity is described  by the relation

\begin{equation}
\begin{aligned}
\langle N(L_X)\;|\;M\rangle = &  K_{norm} \, \langle N_c(L_X)\;|\; M\rangle \times \\
              & ( 1 +  \langle N_{sat}(L_X)\;|\; M\rangle ),
\end{aligned}
\end{equation}

\noindent where $\langle N_{cen}(L_X)\;|\;M\rangle$,  $\langle N_{sat}(L_X)\;|\;M\rangle$  is the mean number of central and satellite AGN respectively, in halos of mass $M$ and with X-ray luminosities brighter than $L_X$. The normalisation term $K_{norm}$ is related to the AGN duty-cycle at fixed X-ray luminosity cut and accounts for the fact that contrary to galaxy samples not all halos above a minimum mass host an active nucleus. The semi-empirical model for populating halos with AGN makes no distinction between central and satellites. As a result the two populations are assumed to have the same duty-cycle and are assigned the same overall normalization. The central and satellite AGN occupations are parametrised as \citep{Zheng2005} 

\begin{equation}\label{eq:hod_central}
\langle N_{cen}(L_X)\;|\;M\rangle = \frac{1}{2} \bigg[ 1 + {\rm erf} \Big( \frac{\log M-\log M_{min}}{\sigma_M} \Big) \bigg],
\end{equation}

\begin{equation}\label{eq:hod_sat}
\langle N_{sat}(L_X)\;|\;M\rangle = 
\begin{cases}
    \bigg( \frac{m-M_0}{M_1} \bigg)^{\alpha}& \text{if } m\geq M_0,\\
    0              & \text{otherwise,}
\end{cases}
\end{equation}

\noindent  where the HOD model parameters are $\log M_{min}$, $\sigma_M$, $\alpha$, $M_0$, $M_1$. The shape of the central HOD is modeled as a softened step function with a cutoff mass $\log M_{min}$. The parameter $\sigma_M$ controls the amplitude of the softening of the step-function profile. The satellite occupation is modeled as a power-law distribution with slope $\alpha$ above the cut-off mass limit $M_0$. The parameter $M_1$ is the normalization of the satellite HOD. Following previous studies \citep[e.g.][]{delaTorre2013} we assume $M_0=M_{min}$ and $M_1 = f_{sat} \, M_{min}$. For galaxy samples $f_{sat}$ is found to be nearly constant and to lie in the range 10-30 \citep[e.g.][]{Zheng2005, Zehavi2011}. Based on these assumptions the AGN HOD model is described by five free parameters, $K_{norm}$, $\log M_{min}$, $\sigma_M$, $\alpha$ and $f_{sat}$. Figure \ref{fig:hod_fit} shows examples of the halo occupation of mock AGN drawn from the MDPL2 simulation boxes and the corresponding parametric fits. It demonstrates that the adopted model parametrisation can adequately describe the halo occupation of mock AGN. Figure \ref{fig:hod_params} presents the best-fit HOD model parameters for mock AGN samples selected from the MDPL2 simulations at different X-ray luminosity cuts and at different redshifts. The results from the SMDPL simulations are consistent with those plotted in Figure \ref{fig:hod_params} and are not shown for the sake of clarity and brevity. There is also overall broad agreement between the HOD model parameters estimated from the mocks that use the \cite{Georgakakis2017_plz} and \cite{Aird2018} specific accretion-rate distributions respectively. The somewhat different behavior of the HOD parameters in the two realizations at the lowest redshift bin ($z=0.25$) of Figure \ref{fig:hod_params} is likely because at these redshifts both the \cite{Georgakakis2017_plz} and the \cite{Aird2018} samples are limited by small numbers. These translate to systematic differences in the corresponding specific accretion-rate distributions that propagate into the AGN mocks and produce the different behavior of the corresponding HOD parameters in  Figure \ref{fig:hod_params}. 

A striking result from Figure \ref{fig:hod_params} is that the HOD parameters, with the exception of the overall normalisation, are not a strong function of luminosity or redshift. There is some variation of these parameters with $L_X$, particularly for the low-redshift sample, but overall these changes are small. At least to the zero-order approximation the mock-AGN halo occupation can be described by a relatively narrow range of parameters for luminosities in the interval $\rm \approx 10^{41}-10^{44}\, erg/s$ and redshifts $z<1.5$. These findings indicate a weak luminosity and redshift-dependence of the AGN clustering, a trend that is also evident in Figure \ref{fig:dmhlx2d}, where the median dark-matter halo-mass is nearly constant with accretion luminosity and redshift. The strong dependence of the HOD normalization parameter on $L_X$ and $z$ in Figure \ref{fig:hod_params} is directly related to the  duty-cycle of the accretion process at fixed redshift and luminosity. 

\begin{figure*}
\begin{center}
\includegraphics[width=2\columnwidth]{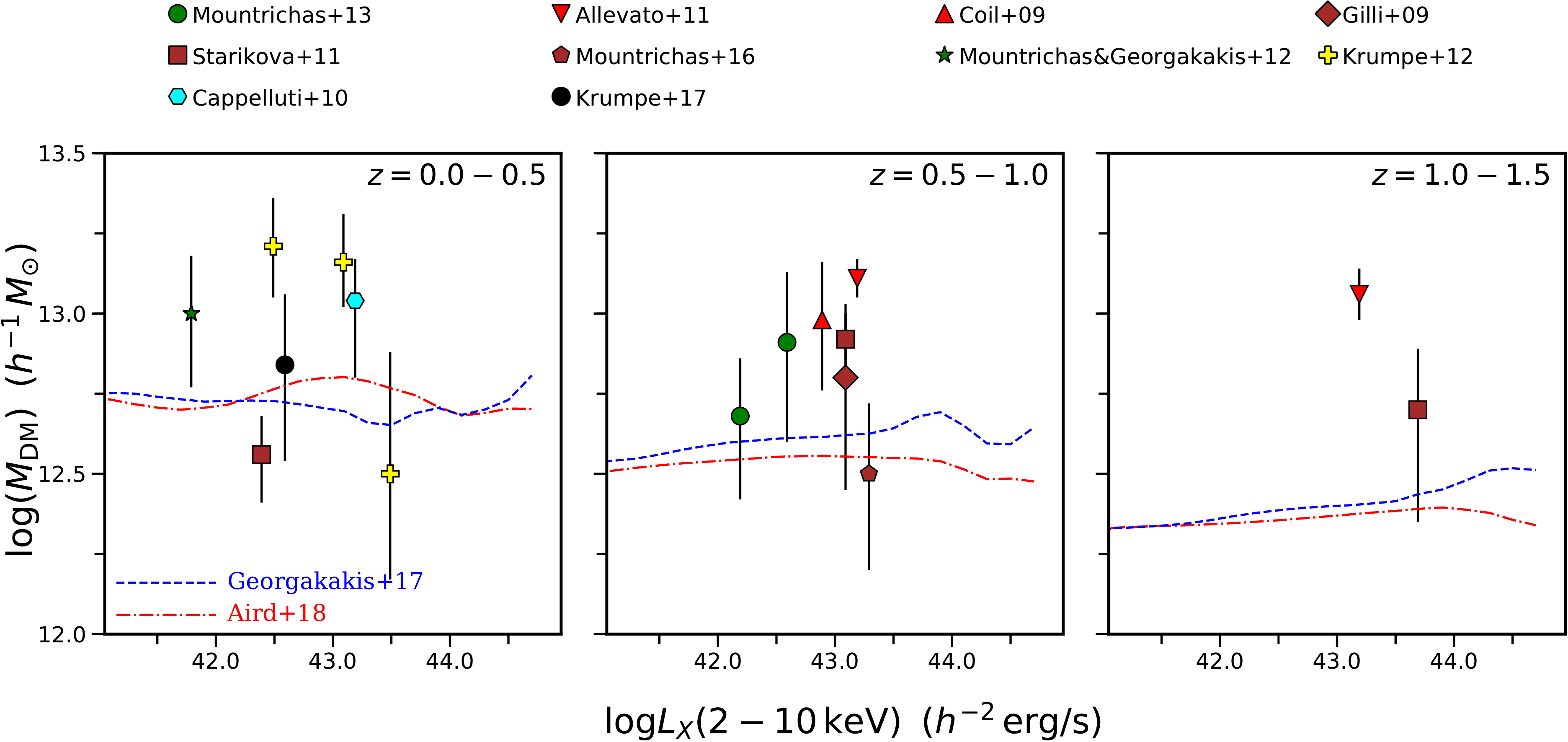}
\end{center}
\caption{Effective dark-matter halo-mass as a function of accretion luminosity for AGN in the MDPL2 mock catalogues constructed using the  \protect\cite{Georgakakis2017_plz} (blue-dashed curves) and \protect\cite{Aird2018} (red dash-dotted curves) specific accretion-rate distributions. The different panels correspond to the simulation boxes at redshifts $z=0.25$, 0.75 and 1.25. The Effective dark-matter halo-mass is estimated from the effective bias (equation \protect\ref{equ:beff}) using the \protect\cite{Comparat2017} parametrisation for the relation between bias and dark-matter halo mass. 
For comparison the mode of the halo mass distribution is expected to be $\log [M_{DM}/M_{\odot}] \approx 12$, i.e. close to the lower limit of the Y-axis in each panel, nearly independent of redshift and luminosity (see Figure \ref{fig:dmhlx2d}).
The data-points are measurements of the mean dark matter halo mass of X-ray selected AGN  at different redshift and luminosity intervals from the literature (see Table \protect\ref{table:xray_samples}). Each data-point is plotted at the panel with redshift closer to mean redshift of the AGN sample from which the measurement is taken. 
}\label{fig:beff}
\end{figure*}

It is also worth pointing out that the parameter $M_{min}$ of the AGN HOD takes values close to $\approx 10^{12} \, M_{\odot}$, thereby indicating that this is the typical environment of accretion events and reiterating earlier conclusions from Figures \ref{fig:dmhlx2d} and \ref{fig:dmhlx1d}. These findings contradict observational studies that estimate mean or typical halo masses for X-ray selected AGN in the range $\log (M_{DM}/M_{\odot})=12.5-13.5$ \citep[e.g.][]{Coil2009, Krumpe2012, Mountrichas2012}. This apparent discrepancy is related to the broadness and skewness of the halo-mass distributions for mock AGN in Figures \ref{fig:dmhlx2d} and \ref{fig:dmhlx1d}. Small number statistics force observers to adopt simpler models for the HOD of X-ray selected AGN or model only the 2-halo term of the correlation function to infer the mean bias of the AGN and the corresponding mean dark matter-halo mass. The latter quantities are offset from the mode or the median in the case of a skewed underlying distribution, like the one in Figures \ref{fig:dmhlx2d} and \ref{fig:dmhlx1d}. This point is demonstrated using the effective bias of mock AGN \citep{Baugh1999}, a quantity that is often used in the literature to approximate the amplitude of the 2-point correlation function that is measured on large scales (2-halo term) by observers. This is defined as

\begin{equation}\label{equ:beff}
b_{\rm eff}(>L_X) = \frac{\int b(M) \, N_{\rm AGN}(>L_X,M)\, n(M)\,{\rm dlog} M }{  \int N_{\rm AGN}(>L_X, M)\,n(M)\,{\rm dlog} M},
\end{equation}

\noindent where  $b(M)$ is the bias at a given halo mass estimated using the parametrisation  presented by \cite{Comparat2017}, $n(M)$ is the number density of dark matter halos of mass $M$, and $N_{\rm AGN}(>L_X,M)$ is the number of AGN that reside in halos of mass $M$ and have 2-10\,keV X-ray luminosity brighter than $L_X$.  Figure \ref{fig:beff} shows the luminosity and redshift dependence of the effective halo mass, i.e that corresponding to the AGN effective bias assuming the model of \cite{Comparat2017} for the conversion. This figure shows that the effective bias of mock AGN corresponds to dark-mater halo masses in the range $\log [M_{DM}/M_{\odot}] \approx 12.25-12.75$, i.e. systematically offset from the mode and median of the mass distributions ($\log [M_{DM}/M_{\odot}] \approx 12$) plotted in Figure  \ref{fig:dmhlx2d}. Also over-plotted in Figure \ref{fig:beff} are observational results on the mean halo-mass of AGN compiled from the literature and presented in Table \ref{table:xray_samples}. The 
model curve is inconsistent with observational measurements that suggest very massive halos for X-ray AGN,  $\log [M_{DM}/M_{\odot}] \ga 13$.  This apparent discrepancy with some of the data-points in Figure \ref{fig:beff} questions the basic assumption of the semi-empirical model, i.e the lack of a physical connection between black-hole accretion events and large-scale environment. It is indeed possible that the diverse fueling/triggering mechanisms proposed in the literature to feed the black holes at the centres of galaxies, e.g. mergers \citep[e.g.][]{DiMatteo2005}, bar-instabilities \citep[e.g.][]{Hopkins_Hernquist2006}, secular evolution \cite[e.g.][]{Ciotti_Ostriker2007}, radio-mode accretion \citep[e.g.][]{Croton2006}, also have an environmental dependence that may imprint detectable signals on the large-scale structure of AGN \citep[e.g.][]{Hopkins2007_LSS, Bonoli2009, Fanidakis2013}. It should also be emphasized however, that the effective bias calculation in Equation \ref{equ:beff} cannot fully capture the non-linear dependence on halo mass of the two-point correlation function statistic. It also does not account for the selection effects of specific samples, e.g. redshift interval, X-ray flux limits. In that respect it is also worth highlighting that some of the discrepant data points in Figure \ref{fig:beff}  correspond to the samples plotted in Figures \ref{fig:wp_agnlrg}-\ref{fig:wp_xxlvipers}, which show good agreement between model and observations. 

\begin{figure}
\begin{center}
\includegraphics[height=0.9\columnwidth]{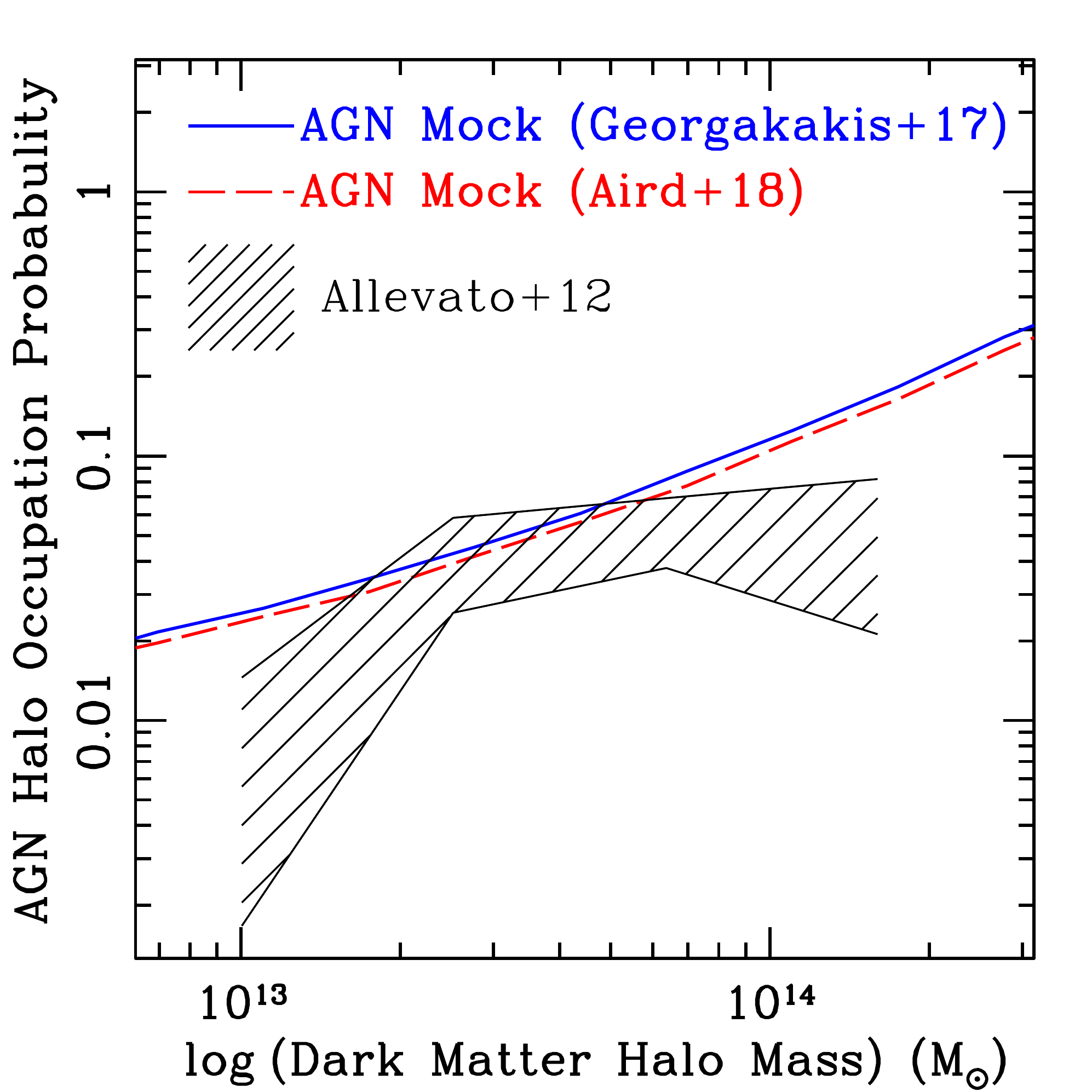}
\end{center}
\caption{The halo occupation of mock AGN in comparison with observations. The grey-hatched region corresponds to the occupation of galaxy groups ($M_{DMH}>10^{13} \,M_{\odot}$) by AGN measured by  \protect\cite{Allevato2012} in the COSMOS field after correcting for luminosity incompleteness and redshift-evolution effects. The blue solid and red long-dashed curves are the predictions of the AGN mocks constructed using the \protect\cite{Georgakakis2017_plz} and the \protect\cite{Aird2018} specific accretion-rate distributions respectively. These curves are calculated from the $z=0.25$ MDPL2 simulation box by selecting AGN  brighter than $L_X(\rm 2-10\, keV )=4\times 10^{42} \rm \ erg \, s^{-1}$.   This cut corresponds to the soft-band (0.5-2\,keV) limit adopted by \protect\citet[$\log L_X(\rm 0.5-2\,keV)/(erg\,s^{-1}) = 42.4$]{Allevato2012} under the assumption of a power-law X-ray spectral energy distribution with $\Gamma=1.9$. 
The mock AGN curve is estimated for $h=0.72$ to allow direct comparison with the observational results of \protect\cite{Allevato2012}.}\label{fig:alle}
\end{figure}

\begin{figure}
\begin{center}
\includegraphics[height=0.9\columnwidth]{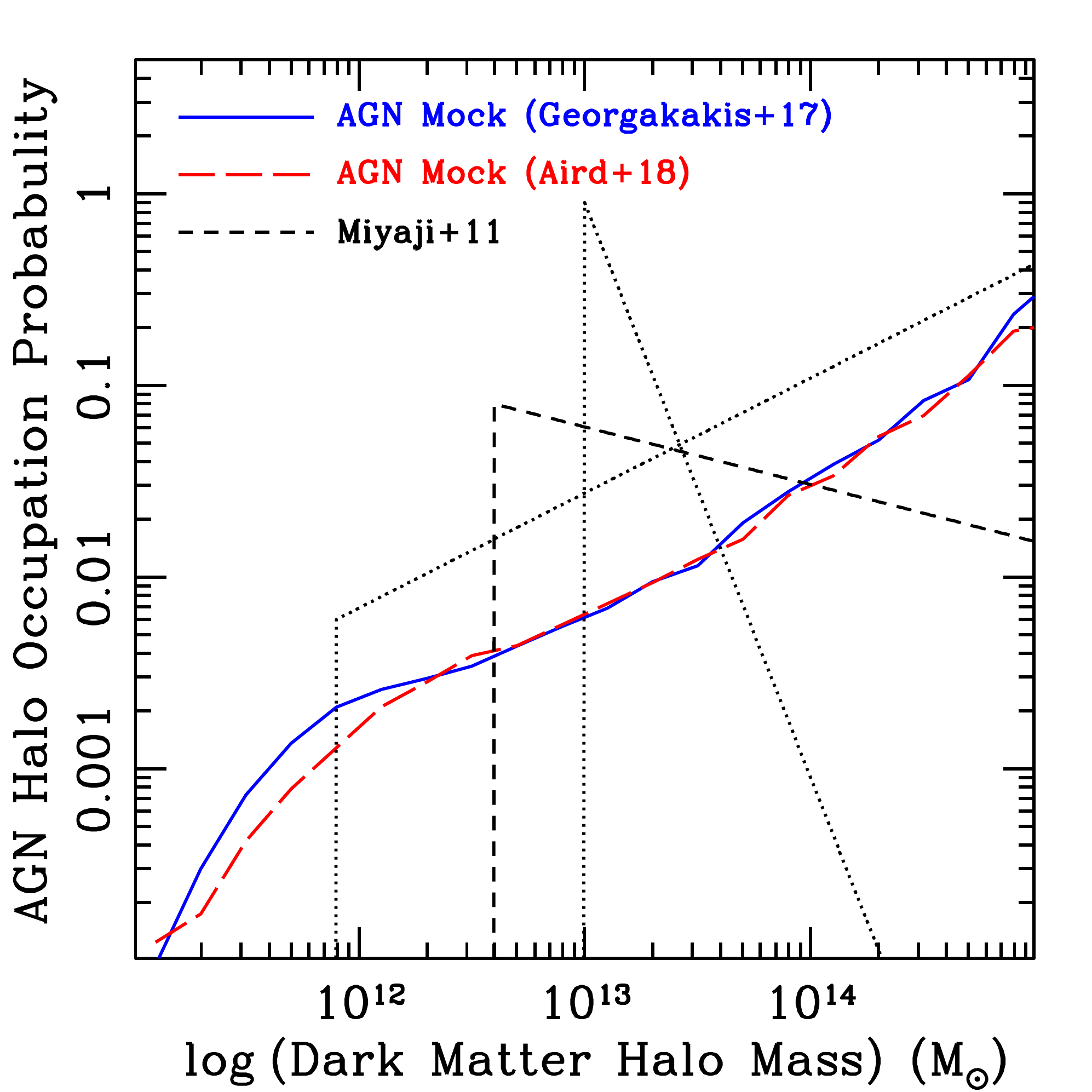}
\end{center}
\caption{The HOD of RASS AGN in the redshift interval $0.16 <z < 0.36$ inferred by \protect\cite{Miyaji2011} assuming a truncated power-law parametrisation. The black sort-dashed line shows the best-fit HOD model. The black-dotted lines represent fits that are consistent with the data within the 68th confidence interval. They provide a measure of the level of uncertainty and the aliases between the model parameters (slope, cutoff halo mass). The blue solid and red long-dashed curves are the predictions of the AGN mocks constructed using the \protect\cite{Georgakakis2017_plz} and the \protect\cite{Aird2018} specific accretion-rate distributions respectively. These curves are derived from the light-cones described in Appendix \ref{appendix:lc_miyaji} that are used to estimate the cross-correlation function of Figure \protect\ref{fig:wp_agnlrg}. The mocks favor relatively steep HOD slopes.}\label{fig:miyajihod}
\end{figure}

We now turn back to Figure \ref{fig:hod_fit} to comment on another feature of the mock-AGN HOD parametric fits, i.e. the prediction for an increasing fraction of satellite AGN toward higher dark matter halo masses. This is quantified by the slope $\alpha$ of the power-law parametrisation of the satellite AGN occupation in Figure  \ref{fig:hod_params}, which is found  in the range $\alpha\approx0.6-1.2$. Such steep slopes are similar to the HODs of galaxy samples and contradict observational results that suggest a flat power-law index $\alpha\la0.6$ for the occupation of AGN in massive halos, $M_{DMH}\ga10^{13}M_{\odot}$ \citep{Miyaji2011, Allevato2012}. The level of discrepancy between model and observations is demonstrated in Figures \ref{fig:alle} and \ref{fig:miyajihod}. \cite{Allevato2012} presented a direct estimate of the AGN occupation in galaxy groups selected at X-ray wavelengths.  Figure \ref{fig:alle} compares their inferred halo occupation distribution with the predictions of the AGN mocks. \cite{Miyaji2011} assumed a truncated power-law HOD model to interpret the cross-correlation function of RASS AGN and SDSS Luminous Red Galaxies. Figure \ref{fig:miyajihod} plots a set of three representative HOD-model fits from the work of \cite{Miyaji2011}, all of which are consistent with the observations at the 68\% confidence level. They visually demonstrate the level of uncertainty in the determination of the AGN HOD, as well as the aliases between model parameters. Also over-plotted in Figure \ref{fig:miyajihod} is the HOD predicted by our semi-empirical model for RASS AGN in the light-cones described in Appendix \ref{appendix:lc_miyaji}. The observational constraints plotted in  Figures \ref{fig:alle}, \ref{fig:miyajihod} are interpreted  by \cite{Miyaji2011} and  \cite{Allevato2012} as evidence that the probability of a halo hosting an AGN is suppressed in high density environments. Such a dependence is not included in the construction of the AGN mocks presented in Section \ref{sec:method} and therefore they do not support flat slopes for the satellite AGN fraction. 

\begin{figure}

\begin{center}
\includegraphics[height=0.9\columnwidth]{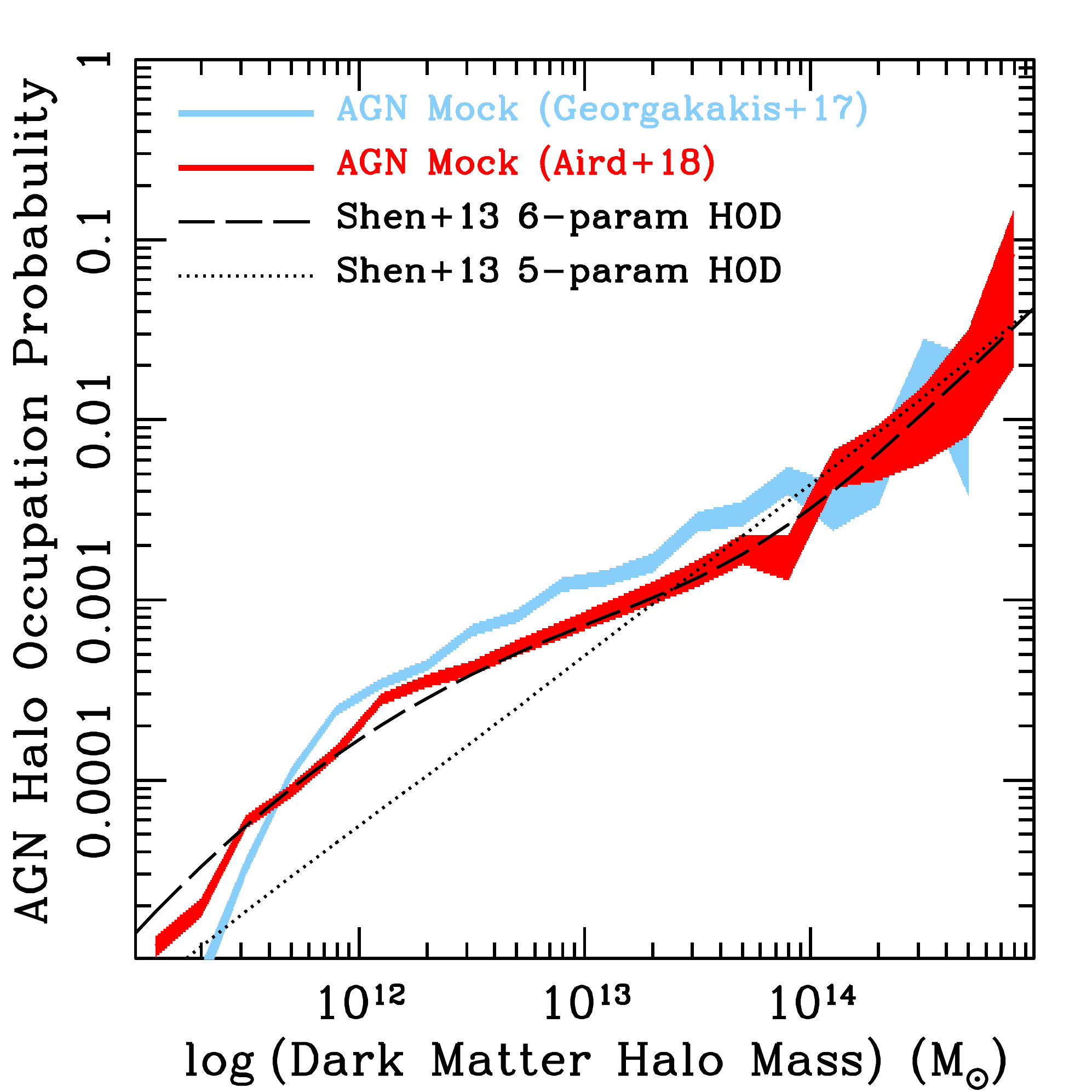}
\end{center}
\caption{The halo occupation of SDSS-DR7 QSOs inferred by \protect\cite{Shen2013}. The black-dotted and black-dashed curves correspond to the best-fit 5-- and 6--parameter HOD models, respectively, adopted by \protect\cite{Shen2013}. The light-blue and red shaded regions show the predictions of the AGN mocks constructed using the \protect\cite{Georgakakis2017_plz} and \protect\cite{Aird2018} specific accretion-rate distributions. These curves are derived from the light-cones described in Appendix \ref{appendix:lc_shen}, which are used to estimate the cross-correlation function of Figure \protect\ref{fig:wp_shen}. The width of shaded regions corresponds to the 68\% uncertainty. This is estimated assuming Poisson statistics for the number of QSOs within a given halo-mass bin. The absolute normalisation of the mock QSO HOD reflects the integrated sky density of mock QSOs and is not important for this comparison, which focus on the shape of the HOD curves. The light-blue and red shaded regions are shifted downward by a factor of two to facilitate the comparison of the shape of these curve with the \protect\cite{Shen2013} HOD-model parametrisations.}\label{fig:shenhod}
\end{figure}

In addition to the X-ray AGN samples discussed above, we also explore the HOD of optically selected QSOs in the semi-analytic simulations and observations. Figure \ref{fig:wp_shen} plots the halo occupation of the SDSS-DR7 QSOs at a mean redshift of $z\approx0.55$ \citep{Shen2013}. This is compared with the expectation from the mocks for an AGN sample that mimics the \cite{Shen2013}  selection function (see Appendix \ref{appendix:lc_shen}). The observationally derived QSO HOD in  Figure \ref{fig:shenhod} has a satellite fraction that increases steeply towards more massive halos, i.e. $\alpha \approx 1$, albeit with large errors ($\approx \pm0.3$).

\begin{table*}
\caption{Compilation of halo mass measurements for X-ray selected AGN samples in the literature.}\label{table:xray_samples}
\begin{center}
\begin{tabular}{ccccc}
	\hline 
	z & z range & $\log M_{\rm DM}$ & $\log L_X(\rm 2-10\,keV)$ & Reference \\	 
     & & ($h^{-1}\,M_{\odot}$) & ($h^{-2} \rm \, erg \, s^{-1}$) & \\
 (1) & (2) & (3) & (4) & (5) \\
\hline 
	0.05 & 0.00-0.15 & 13.20$^{+0.13}_{-0.24}$ & 43.2 & \protect\cite{Cappelluti2010} \\[3pt]
    0.10 & 0.03-0.20 & 13.00$^{+0.18}_{-0.23}$ & 41.8 & \protect\cite{Mountrichas2012} \\[3pt]
	0.69 & 0.40-0.90 & 12.68$^{+0.18}_{-0.26}$ & 42.2 & \protect\cite{Mountrichas2013} \\[3pt]
	0.97 & 0.70-1.40 & 12.91$^{+0.22}_{-0.31}$ & 42.6  & \protect\cite{Mountrichas2013} \\[3pt]
	0.81 & 0.50-1.20 & 12.50$^{+0.22}_{-0.30}$ & 43.3  & \protect\cite{Mountrichas2016} \\[3pt]
0.02 & 0.01-0.04 & 12.84$^{+0.22}_{-0.30}$ & 42.6 & \protect\cite{Krumpe2017} \\[3pt]
  0.13 & 0.07-0.16 & 13.21$^{+0.15}_{-0.16}$ & 42.5 & \protect\cite{Krumpe2012} \\[3pt]
	0.27 & 0.16-0.36 & 13.16$^{+0.15}_{-0.14}$ & 43.1 & \protect\cite{Krumpe2012} \\ [3pt]
	0.42 & 0.36-0.50 & 12.50$^{+0.38}_{-0.33}$ & 43.5 & \protect\cite{Krumpe2012} \\[3pt]
	0.80 & - & 13.11$^{+0.06}_{-0.06}$ & 43.2 & \protect\cite{Allevato2011} \\[3pt]
    1.30 & - & 13.06$^{+0.08}_{-0.08}$ & 43.2 & \protect\cite{Allevato2011} \\[3pt]
	0.90 & 0.70-1.40 & 12.98$^{+0.18}_{-0.22}$ & 42.9 & \protect\cite{Coil2009} \\[3pt]
 	0.94 & 0.40-1.60 & 12.80$^{+0.20}_{-0.35}$ & 43.1 & \protect\cite{Gilli2009} \\[3pt]
	0.37 & 0.17-0.55 & 12.56$^{+0.12}_{-0.15}$ & 42.4 & \protect\cite{Starikova2011} \\[3pt]
	0.74 & 0.55-1.00 & 12.92$^{+0.11}_{-0.15}$ & 43.1 & \protect\cite{Starikova2011} \\[3pt]
	1.28 & 1.00-1.63 & 12.70$^{+0.19}_{-0.35}$ & 43.7 & \protect\cite{Starikova2011}\\[3pt]	
\hline	 	 	 	 
\end{tabular}
\end{center}
\begin{list}{}{}
\item \noindent (1) The median redshift of the AGN sample; (2) the redshift range of the sample, if available in the relevant publication; (3) the typical dark matter halo mass of the AGN sample estimated via correlation function analysis. The units are $h^{-1}\,M_{\odot}$; (4) the average $2-10\,\rm{keV}$ X-ray luminosity of the AGN sample in units of $h^{-2}\rm \, erg \, s^{-1}$. For samples for which the AGN X-ray luminosity is estimated in an energy interval other than the 2-10\,keV band we convert to $L_X(\rm 2-10\,keV)$ assuming a power-law X-ray spectrum with photon index of $\Gamma=1.9$; (5) reference to the relevant paper for each AGN sample.
\end{list}
\end{table*}


\section{Discussion}

We propose an empirical model to populate the dark matter halos of cosmological simulations with active black holes and produce mock catalogues of AGN. Our method is based on empirical relations that associate dark matter halos with galaxy stellar masses. This information is further combined with observationally determined AGN specific accretion-rate distributions to quantify the probability of a galaxy hosting an accretion event. An assumption of our approach is that the probability of an accretion event onto supermassive black holes  does not depend on environment, i.e. the mass of the dark matter halo that hosts an AGN. Under this assumption the large-scale distribution of AGN is essentially that of galaxies modulated by the specific accretion-rate distribution of active black holes. The resulting mock catalogues reproduce by construction the observed X-ray luminosity function of AGN and its redshift evolution, the stellar mass function of galaxies and the halo vs stellar-mass relation of the galaxies.

Using the semi-empirical model above we populate the MultiDark cosmological simulation with AGN and show that their clustering properties, measured by the two-point correlation-function statistic, are consistent with state-of-the-art observational measurements of X-ray or UV/optically selected samples at different redshifts and accretion luminosities (Figures \ref{fig:wp_agnlrg}-\ref{fig:wp_shen}). This agreement, within the error budget of the current observations and simulations, supports the model assumption that the AGN activity (at least in central halo galaxies) and the large-scale environment are unrelated. It also gives us confidence that the semi-empirical model provides a realistic representation of the large-scale distribution of AGN and hence, can be used to explore and  draw conclusions on the mass distribution of their dark-matter halos. 

A feature of the semi-empirical model is that the mock AGN are hosted by dark matter halos with a relatively wide range of masses. In Figure \ref{fig:dmhlx1d} the peak of the distribution is at $M_{DM} \approx 10^{12} \, \rm M_{\odot}$ with a tail extending to $M_{DM} \ga 10^{13} \, \rm M_{\odot}$. This broadness is also mirrored in the HOD of central AGN, which can be described by an error function (softened step-function, see Equation \ref{eq:hod_central}) with a turnover-mass parameter $\log M_{min}/M_{\odot} \approx 12$ (Figure \ref{fig:hod_params}). These properties of the mock AGN in our semi-empirical simulations are broadly consistent with observational constraints of the HOD of X-ray AGN \citep{Miyaji2011, Allevato2012} and optically selected QSOs \citep[e.g.][]{Richardson2012, Shen2013_cl}. A broad distribution of halo masses for moderate-luminosity X-ray AGN at $z<1$ is also proposed by \cite{Leauthaud2015}. They show that the weak-lensing signal of this population is consistent with the assumption that their hosts are drawn from the normal (non-AGN) galaxy population. They then combine the observed stellar-mass and redshift distribution of the AGN sample with the halo-stellar mass relation of normal galaxies \citep{Leauthaud2012} to infer a halo mass distribution for the AGN that is similar to Figure \ref{fig:dmhlx2d}. A consequence of the skewness of the AGN halo-mass distribution is that measurements of the typical or mean dark-matter halo mass, inferred by e.g. modeling only the one-halo term of the correlation function, are biased to high values compared to the mode (see Figure \ref{fig:beff}). Observational evidence that the bulk of the X-ray AGN population is associated with moderate size halos and a tail extending to high masses has been presented  by \cite{Mountrichas2013}. They showed  that the typical/mean halo mass of moderate-luminosity X-ray AGN at $z\approx1$ decreases by 0.5\,dex, from  $\log [M_{DM}/\rm h^{-1} M_{\odot}] \approx 13.2$ to  $\sim 12.7$, once a small fraction of sources  (5\%) associated with galaxy groups ($M_{DM} \ga 2\times 10^{13}\, \rm M_{\odot}$) is removed from the sample. 

Observational studies have also investigated the luminosity dependence of AGN clustering with mixed results \citep[e.g.][]{Coil2009, Krumpe2012, Koutoulidis2013, Fanidakis2013, Shen2013_cl, Mountrichas2016}. Our semi-empirical model predicts no, or at best a very weak dependence of the AGN clustering on the accretion luminosity. This is very different from galaxy samples, for which there is a well established correlation between dark-matter halo mass and the luminosity of the stars \cite[e.g.][]{Zehavi2005}. This is because stellar luminosity, particularly at longer wavelengths, correlates with the stellar mass of galaxies. In contrast the AGN accretion luminosities trace only loosely and indirectly the stellar mass of their host galaxies. The  quasi power-law form of the observationally-derived specific accretion-rate distributions with a steep decrease towards high specific accretion-rates \citep{Aird2012, Bongiorno2012, Bongiorno2016, Georgakakis2017_plz, Aird2018} means that AGN at any luminosity cut are preferentially associated with galaxies close to knee of the stellar-mass function \citep{Aird2013}, i.e. stellar masses $M_{star} \approx 10^{10}-10^{11}\,M_{\odot}$ \citep{Georgakakis2017_plz}. This stellar-mass range roughly corresponds to the position of the break of the dark-matter vs stellar mass relation, which occurs at $M_{DM}\approx10^{12}\,h^{-1} \, M_\odot$ over a broad range of redshifts \citep[e.g.][]{Behroozi2013}. AGN are therefore expected to be associated with dark matter halos with median mass $M_{DM}\approx10^{12}\,h^{-1} \, M_\odot$ nearly independent of accreion luminosity.

\begin{figure*}
\begin{center}
\includegraphics[width=2\columnwidth]{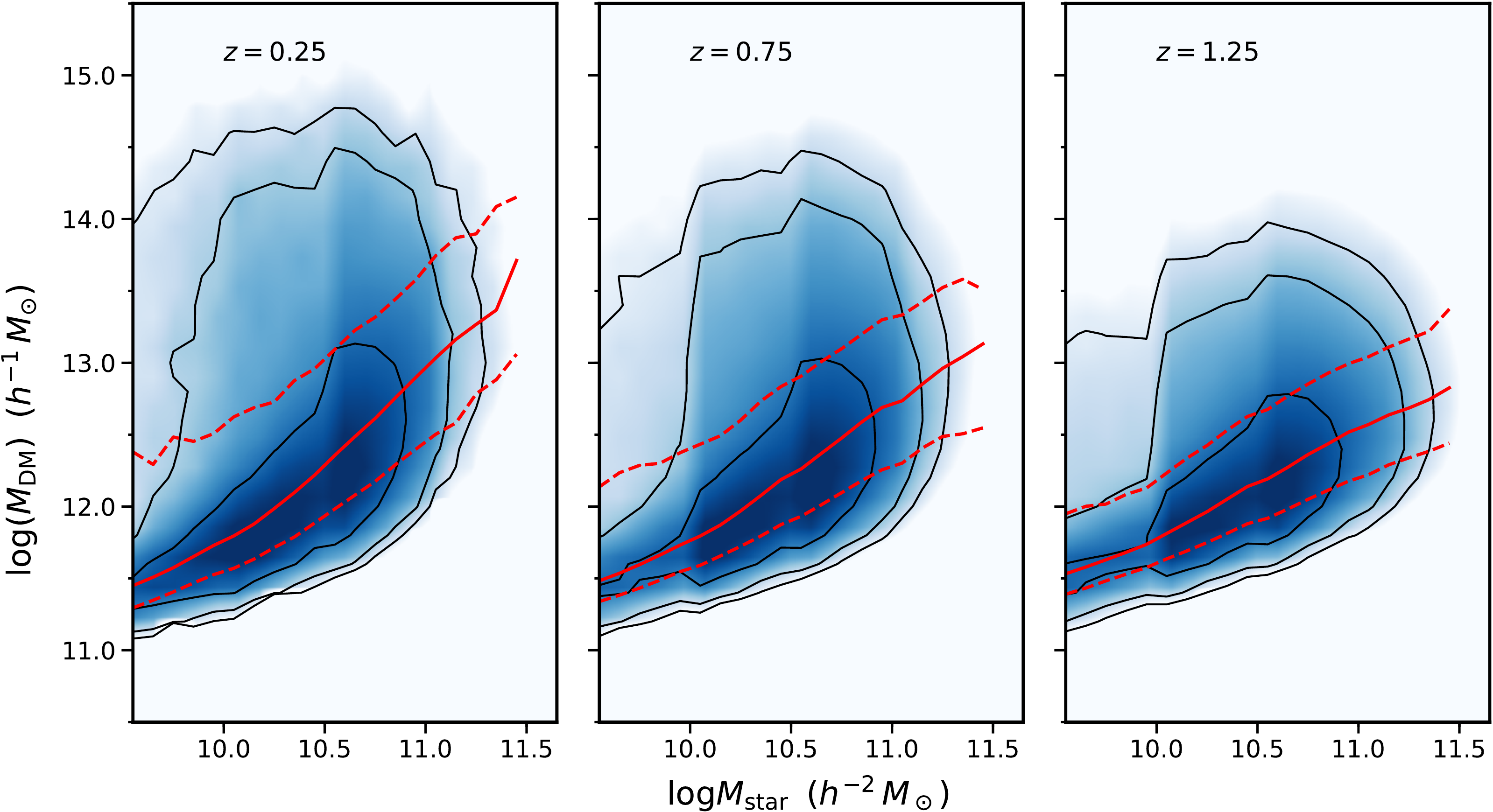}
\end{center}
\caption{Mass distribution of the dark-matter halos that host AGN as a function of the stellar mass of the underlying galaxy. AGN are selected from the MDPL2 simulation boxes with accretion luminosity brighter than $L_X>\rm 10^{42} \, erg \, s^{-1}$. The panels correspond to the redshifts of the simulation boxes used in this work, $z=0.25$, 0.75 and 1.25. The contour levels are chosen to enclose 68, 95 and 99\% of the total number of mock AGN in the simulation box. The red solid line is the median of the distribution at fixed stellar mass. The dotted lines mark the 16th and 84th percentiles of the distribution. The specific-accretion rate distributions of \protect\cite{Aird2018} are used to populate halos with AGN in this plot. The observed trends remain unchanged if instead we use the \protect\cite{Georgakakis2017_plz} specific-accretion rate distributions or if we adopt luminosity-cuts other than $L_X>\rm 10^{42} \, erg \, s^{-1}$ to select mock AGN.
}\label{fig:mass2D}
\end{figure*}

Contrary to the accretion luminosity, it is the AGN host-galaxy stellar mass that is inherently correlated to halo mass in our semi-empirical methodology. This is shown in Figure \ref{fig:mass2D}, which plots stellar vs halo mass at fixed accretion luminosity. The mean halo-mass of the distribution in that figure increases toward higher stellar mass. This property of the model is consistent with recent observational evidence for a dependence of the AGN clustering on the properties of their hosts, such as the stellar mass. \cite{Georgakakis2014_cluster} for example, found that X-ray AGN have similar clustering properties to non-AGN samples of star-forming and quiescent galaxies selected to have similar stellar mass distributions to the AGN hosts. This finding suggests that the level of clustering of X-ray selected AGN samples primarily correlates with the stellar masses of their host galaxies, rather than their instantaneous accretion luminosities. \cite{Mendez2016} showed that the correlation function of infrared, X-ray and radio-selected AGN is similar to that of control non-active galaxy samples matched in stellar mass, star-formation rate and redshift to the AGN. This result further emphasizes the importance of covariances between galaxy properties and the large-scale environment of active black-hole samples selected at different wavebands.

A property of the halo occupation of the AGN that may differentiate them from galaxies, and hence point to black-hole fueling physics, is the halo-mass dependence of the satellite fraction. For galaxy samples the fraction of satellites increases with increasing halo mass. This trend can be parametrised by a power-law distribution with index that typically takes values $\alpha\approx1$ \cite[e.g.][]{Zehavi2005, Zehavi2011}. This may be in tension with some recent observational results that suggest a flatter slope for the AGN population \citep{Miyaji2011, Allevato2012}, possibly related to the suppression of accretion events in dense environments \citep[e.g.][]{Kauffmann2004, Popesso_Biviano2006}. The latter scenario is debated however, e.g. by observations that estimate similar fractions of X-ray selected AGN in clusters \citep{Martini2007} and the field \citep{Haggard2010}, or clustering studies of optically-selected QSOs that find satellite occupations consistent with a power-law slope of $\alpha\approx1$ \citep{Richardson2012,Shen2013_cl}. The semi-empirical model presented here does not include any halo-mass dependent terms for the galaxy occupation of AGN. It is therefore not surprising that the satellite halo-occupation of mock AGN is described by a steep power-law model index, $\alpha\approx1$, comparable to galaxy samples (see Fig.\ref{fig:hod_params}). Nevertheless, the comparison of the model predictions with the observationally determined HODs of in Figures \ref{fig:alle}, \ref{fig:miyajihod} suggests that the level of tension is still small. Improving current constraints on the halo occupation of satellite AGN requires larger samples, particularly at X-rays, to reduce shot-noise in measurements of the 1-halo term of the AGN two-point correlation function, or estimates of the AGN incidence in groups and clusters of galaxies. The eROSITA All Sky Survey will yield a large and homogeneous sample of X-ray selected AGN \citep[$\approx4\times10^{6}$,][]{Merloni2012} and has the potential to constrain the 1-halo term behavior of this population. Mock catalogues of the eROSITA X-ray sky and predictions on the expected clustering  signal of eROSITA AGN will be presented in a future paper (Comparat et al. in prep.).

Overall the analysis presented in this work underlines the role of AGN host-galaxy properties, such as stellar mass, for understanding the observed clustering properties of samples of active supermassive black-holes. Disentangling the impact of galaxies on the observed signal is key for interpreting any residuals in the context of black-hole fueling physics and AGN triggering mechanisms.

\section{Summary and conclusions}

A semi-empirical model for the distribution of AGN on the cosmic-web is developed by combining cosmological dark-matter halo simulations with observational estimates of the incidence of accretion events in galaxies. We first show that the model is consistent with current measurements of the two-point correlation function of X-ray AGN samples and then use it to explore the halo-mass distribution of active black holes. The main features of the model are 

\begin{itemize}

\item Mock AGN are hosted by dark matter halos with a broad range of masses. The mode of the mass distribution lies at $M_{DM} \approx 10^{12} \, \rm M_{\odot}$ with a tail extending to cluster-size halos. 

\item The fundamental properties of the mock-AGN halo mass distribution are nearly independent of accretion luminosity and redshift. This translates to a weak luminosity and redshift dependence of the AGN clustering at least to $z<1.5$ and for luminosities in the interval $L_X(\rm 2-10\,keV) \approx 10^{41} - 10^{44} \, erg/s$. 

\item The halo occupation of the mock AGN at different accretion luminosity cuts can be described by the 5-parameter halo model of \cite{Zheng2005} that is widely used for galaxy samples. 

\item the incidence of AGN in the central galaxies of halos is independent of the halo mass.   

\item The AGN satellite fraction in the model increases with increasing halo mass, in a manner similar to galaxy samples. This contradicts some observational studies that suggest that the AGN occupation of satellites is flatter than that of galaxies.

\end{itemize}

\section{Acknowledgements}

JA acknowledges support from an STFC Ernest Rutherford Fellowship, grant
code: ST/P004172/1. The CosmoSim database used in this paper is a service by the Leibniz-Institute for Astrophysics Potsdam (AIP). The MultiDark database was developed in cooperation with the Spanish MultiDark Consolider Project CSD2009-00064. The authors gratefully acknowledge the Gauss Centre for Supercomputing e.V. (www.gauss-centre.eu) and the Partnership for Advanced Supercomputing in Europe (PRACE, www.prace-ri.eu) for funding the MultiDark simulation project by providing computing time on the GCS Supercomputer SuperMUC at Leibniz Supercomputing Centre (LRZ, www.lrz.de).




\bibliographystyle{mnras}
\bibliography{mybib} 
\appendix
\section{light cone for LRG and RASS AGN}\label{appendix:lc_miyaji}

In this section we describe the construction of the light cone that includes SDSS Luminous Red Galaxies \citep[LRGs,][]{Eisenstein2001} and ROSAT All-Sky Survey \citep[RASS,][]{Voges1999} AGN in the redshift interval $z=0.16-0.36$.  The resulting light cone is used to estimate the LRG/RASS-AGN cross-correlation function and compare with the observational results of \cite{Miyaji2011}. 

For this application we use the MDPL2 simulation box at redshift $z=0.25$. The motivation for the specific snapshot redshift is because it lies close to the middle of the redshift interval of interest, $z=0.16-0.36$. The halos of the simulation are populated with LRGs using the 5-parameter HOD parametrisation of \citet[][their Appendix B]{Zheng_Zehavi2009} for galaxies with $g$-band absolute magnitudes $M_g<-21.2$\,mag. Central halos of the simulation are assigned LRGs based on the central-galaxy halo-occupation probability given by the \cite{Zheng_Zehavi2009} model. The number of satellite LRGs of a halo with a central LRG is assumed to follow a Poisson distribution with expectation value estimated from the HOD model. LRGs are then randomly assigned to the satellites of the central halo identified by the {\sc Rockstar} finder \citep{Behroozi2013_rock}. It is noted that the LRGs are associated with massive halos, $\ga 10^{13}\,h^{-1} M_{\odot}$, and therefore the resolution of the MDPL2 simulation (particle mass $1.5\times10^9 \, h^{-1} \, M_{\odot}$) is sufficient to reproduce their clustering properties. Dark-matter halos are further populated with AGN following the methodology described in Section \ref{sec:method}. 

The simulation box is then projected onto the sky by placing the observer at one of the corners. The resulting light cone covers 1/8 of the sphere. X-ray luminosities in the $2-10$\,keV band are converted to fluxes in the ROSAT $0.1-2.4$\,keV energy interval assuming a power-law X-ray spectrum with index $\Gamma=1.9$. The flux cut $f_X(\rm 0.1-2.4\,keV)=10^{-13}\, erg \, s^{-1} \, cm^{-2}$, is then applied to the mock catalogue to produce an AGN sample that mimics the RASS selection.  The projected cross-correlation function between mock LRGs and RASS AGN in the redshift interval $0.16<z<0.36$ is  estimated by integrating along the line-of-sight direction to scales $\pi_{max}=\rm 80\,h^{-1}\,Mpc$, i.e. the projection depth adopted by \cite{Miyaji2011}.

\section{Light-Cone for VIPERS galaxies and XMM-XXL AGN}\label{appendix:lc_vipers}

Cosmological dark-matter halo simulations are used to construct the light-cone that mimics the selection of the VIPERS \citep{Guzzo2014} galaxies and the XMM-XXL AGN \citep{Liu2016, Pierre2016}. This product is then used to estimate the AGN/galaxy cross-correlation function in the redshift interval $z\approx 0.6-1.0$ and compare with the observational results of \cite{Mountrichas2016}. 

For this exercise we choose to use the larger MDPL2 simulation at $z=0.75$ to produce a light-cone in the redshift range $z\approx 0.6-1.0$ using a single snapshot and avoid box repetition. We limit the analysis to halos with masses $\ga 10^{11}\,h^{-1} \rm \, M_\odot$, which contain at least $\approx 70$ dark matter particles.  This mass limit is sufficient to study the central-galaxy component of the VIPERS HOD \citep{delaTorre2013}. We caution that satellite VIPERS galaxies may be associated with halos below this limit. For such systems the MDPL2 simulation is therefore affected by incompleteness. Nevertheless, the VIPERS-galaxy/AGN cross-correlation function of \cite{Mountrichas2016} is limited to scales $\ga 1\,\rm Mpc$, where the 2-halo term dominates the signal. For our purposes it is therefore sufficient to populate only central halos with VIPERS galaxies and ignore the satellite contribution to the cross-correlation function. 

The 5-parameter HOD model presented by \cite{delaTorre2013} is used to populate central halos of the MDPL2 simulation with VIPERS galaxies. We first project the simulation on the sphere by aligning the line-of-sight direction along the $Z$-axis of the simulation box. The centre of the box is offset along the $Z$-axis by the comoving distance that corresponds to the redshift $z=0.75$. The observer is placed at comoving coordinates $(X,Y,Z)=(500, 500, 0)\,h^{-1}\rm \,Mpc$, i.e. at redshift $z=0$.  This setup produces a light-cone that extends from redshift $z\approx0.51$ to $z\approx1.03$ without box repetition and provides a field of view of about $\rm 450 \, deg^2$ (12\,deg radius). Halos at different redshifts are populated with VIPERS galaxies using the parametrisation of \cite{delaTorre2013}  for galaxies brighter than the $B$-band absolute magnitude $M_B-5\,\log(h) =-19.0\rm \,mag$. Following the methodology of \cite{Skibba2007} central galaxies are assigned an $M_B$ based on the relation between the minimum halo-mass parameter of the HOD model, $M_{min}$, and the $B$-band absolute magnitude given by \cite{delaTorre2013}. Absolute magnitudes are then converted to observed $i$-band fluxes assuming the Spectral Energy Distribution of the Sb-type galaxies of \cite{Ilbert2009}. The assumption for a single SED for the estimation of k-corrections is for simplicity. The choice of the Sb type is because it is intermediate between passive and star-forming galaxies. It is also worth emphasizing that at $z\approx0.8$ the rest-frame wavelength of the $B$-band is close to the observers-frame $i$-band wavelength and hence, the k-correction for the conversion of the $M_B$ absolute magnitudes to $i$-band fluxes is small. The resulting galaxy catalogue  is thresholded at $i<22.5$\,mag to mimic the $i$-band target selection of the VIPERS.

The halos of the MDPL2 simulation are populated with AGN following the steps of Section \ref{sec:method}. X-ray fluxes in the 0.5-2\,keV band are estimated from the 2-10\,keV luminosities assuming a power-law X-ray spectrum with index $\Gamma=1.9$. The fluxes are then filtered through the XMM-XXL 0.5-2\,keV sensitivity curve \citep{Liu2016} to mimic the selection of the sample used by \cite{Mountrichas2016}. The X-ray sensitivity curve measures the probability that a source with a certain flux is detected within the surveyed area. This curve is first normalised to unity. Then, for each mock AGN with flux $f_X(\rm 0.5-2\,keV)$ a random number is generated in the interval $0-1$. If the value of the normalised sensitivity curve at the source flux is larger than the random number, then the source is retained in the mock catalogue. 

\begin{figure}
\begin{center}
\includegraphics[height=0.9\columnwidth]{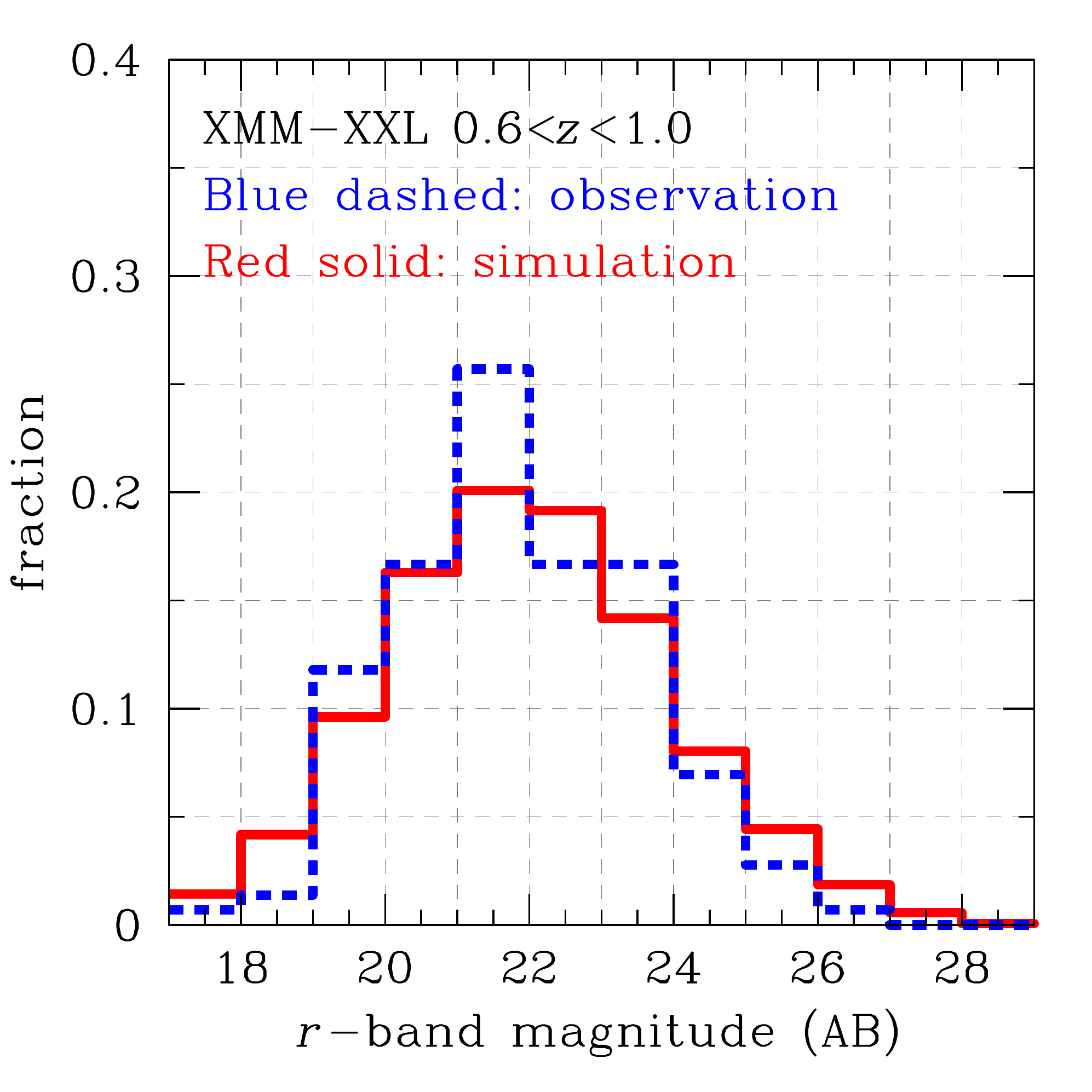}
\end{center}
\caption{Comparison of the optical ($r$-band) magnitude distribution of the XMM-XXL AGN in real observations and the simulation. The red-dashed line is the XMM-XXL sample presented by \protect\cite{Georgakakis2017XXL}. Only sources in that catalogue with spectroscopic or photometric redshifts in the interval $z=0.6-1.0$ are plotted. The solid-blue line is the simulation prediction after applying the same X-ray selection as in the real data. Optical fluxes for mock AGN are estimated following the methodology described in the text.}\label{fig:XXLnz}
\end{figure}

The sample used by \cite{Mountrichas2016} is also limited to the optical magnitude $r\approx22.5$\,mag, because of the requirement for spectroscopic redshift measurements, the majority of which is from the SDSS \citep{Menzel2016}. We account for this selection by assigning optical fluxes to individual sources. They are estimated as the superposition of AGN light and stellar emission from the host galaxy. For this exercise assumptions are made on the intrinsic SED of active black holes, the fraction of obscured AGN and the star-formation history of galaxies that host an active nucleus. We defer a detailed description of the methodology of assigning optical fluxes to mock AGN to a future paper (Georgakakis et al. in prep.) and outline here the most important steps and assumptions. We first assume that the obscured AGN fraction is a function of luminosity and redshift and follows the parametrisation of \cite{Merloni2014}. Obscured AGN do not contribute to the optical part of the spectrum, i.e. $r$-band. For such sources the observed optical fluxes are from the host galaxy only. For unobscured AGN we assume the type-I QSO template SED of \cite{Salvato2011}, which is normalised so that the luminosity densities at 2\,keV and 2500\AA \, follow the $L_{\nu}({\rm 2keV})-L_{\nu}({\rm 2500\mathring{A} })$ correlation of \cite{Lusso2010} with an intrinsic scatter of 0.2\,dex. Star-forming AGN hosts with a given stellar mass are placed on the Main Sequence of star-formation \citep{Schreiber2015} with a scatter of 0.2\,dex. Passive galaxies are assigned star-formation rates 2\,dex below the main sequence. A delayed star-formation history model is adopted to synthesise the stellar population of a galaxy with a given stellar mass, star-formation rate and redshift, and hence determine its SED and observed optical magnitudes. For the latter calculation we use the {\sc cigale} code \citep[Code Investigating GALaxy Emission; ][]{Noll2009, Ciesla2015}. Figure \ref{fig:XXLnz} demonstrates the performance of the simple approach outlined above in reconstructing the optical fluxes of the AGN population. This figure 
compares the observed $r$-band magnitude distribution of the sample of the XMM-XXL AGN presented by \cite{Georgakakis2017XXL} in the photometric or spectroscopic redshift interval $0.6<z<1.0$ with that predicted from the simulation. The magnitude limit $r<22.5$\,mag  is applied to the XXM-XXL AGN light cone to mimic the spectroscopic follow-up selection of the \cite{Mountrichas2016} sample.

\section{Light-Cone for the SDSS Main galaxy sample}\label{appendix:lc_sdssmain}

We construct a light cone that represents the SDSS Main Galaxy sample \cite{Strauss2002}. This is cross-correlated with mock AGN that follow the selection functions of the RASS \citep{Voges1999} and the XMM/SDSS \citep{Georgakakis_Nandra2011} surveys. The resulting AGN/galaxy projected correlation functions are compared with the observational results of \citet{Krumpe2012} and \citet{Mountrichas2012}, respectively. These studies measure the AGN/galaxy cross-correlation function to scales below $1\,h^{-1}\,\rm Mpc$. For this comparison it is therefore important to include satellite galaxies in the light cone.  To this end, we use the higher resolution SMDPL simulation with a box size of $400\,h^{-1}\,\rm Mpc$ and limit the analysis to halos with masses $\ga 3\times10^{10}\,h^{-1} \rm \, M_\odot$. These are populated with galaxies using the 5-parameter HOD model of \cite{Zehavi2011}. The methodology of \cite{Skibba2007} is adopted to assign $r$-band absolute magnitudes, $M_r$, to central and satellite halos. 

For the light-cone construction the observer is placed at the corner of the simulation box to produce a field-of-view of $\pi/2$ steradians (1/8 of the of sphere surface area). The SMDPL boxes are repeated to produce a cube with a side of $800\,h^{-1}\,\rm Mpc$. This is sufficient to  produce AGN and galaxy mocks to redshift $z\approx0.28$. The Absolute magnitudes, $M_r$,  are converted to observed $r$-band fluxes assuming the Spectral Energy Distribution of the Sb-type galaxies of \cite{Ilbert2009}. The resulting galaxy catalogue  is thresholded to $r<17.6$\,mag to mimic the target selection of the SDSS Main galaxy sample \citep{Strauss2002}. For the comparison with the results of \cite{Krumpe2012} we apply the absolute magnitude and redshift cuts that correspond to their low-redshift sample, $-21<M_r-5\,\log(h) <-20.0\rm \,mag$ and $0.07<z<0.16$, respectively.  \citet{Mountrichas2012} use the SDSS Main galaxies in the redshift interval $0.02<z<0.2$. The same selection is also applied to the mock catalogue. 

Halos are populated with AGN as explained in Section \ref{sec:method}. We mimic the RASS sample selection of \cite{Krumpe2012} by applying a flux limit of $f_X(\rm 0.1 - 2.4 \, keV) > 10^{-13}\, erg \, s^{-1}$. The 0.5-10\,keV X-ray sensitivity curve of the XMM/SDSS survey is used to filter the mock AGN and reproduce the X-ray sample selection of \cite{Mountrichas2012}. In both cases the X-ray fluxes are estimated from the 2-10\,keV luminosities assuming a power-law X-ray spectrum with index $\Gamma=1.9$.

\section{Light-Cone for the SDSS QSOs and CMASS galaxies}\label{appendix:lc_shen}.

This section describes the construction of light-cones that resemble the AGN and galaxy samples used by \cite{Shen2013}. They estimated the cross-correlation function between SDSS-DR7 QSOs \citep{Schneider2010}  and the CMASS (constant mass) galaxies \citep{Nuza2013} of the SDSS-III/BOSS \cite[Baryon Oscillation Spectroscopic Survey][]{Dawson2013} survey. The CMASS selection is designed to yield luminous red galaxies at redshifts  $z\ga 0.4$ to a limiting stellar mass of $\approx 10^{11}\,M_{\odot}$. Our analysis focuses on the redshift interval $0.43 < z < 0.7$, where the bulk of the CMASS galaxies lies \citep{Nuza2013}. This is narrower than the redshift range adopted by \cite[][$z=0.3-0.9$]{Shen2013}. The number density of the CMASS galaxies however, drops substantially at $z\la0.43$ or $z>0.7$ \citep[e.g.][]{White2011, Alam2017} and hence, these redshift intervals have a minor contribution to the QSO/CMASS cross-correlation signal. The SDSS-DR7 QSO sample consists of all quasars targeted as part of the the SDSS-I/II spectroscopic surveys. For the redshift interval of interest it is essentially magnitude limited to $i<19.1$\,mag. 

The MDPL2 simulation box at a  snapshot redshift of $z=0.55$ is projected to the sky. The adopted redshift corresponds to the mean of the redshift distribution of the CMASS galaxies. The centre of the box is offset along the $Z$-axis by the comoving distance of the redshift $z=0.55$. The observer is placed at comoving coordinates $(X,Y,Z)=(500, 500, 0)\,h^{-1}\rm \,Mpc$, i.e. at redshift $z=0$.  This setup produces a light-cone that extends from redshift $z\approx0.34$ to $z\approx0.79$ without box repetition and provides a field of view of about $\rm 706 \, deg^2$ ($\approx 15$\,deg radius). The sky area of the light-cone is smaller than that of the real observations used by \citet[][$\rm 6248s\,deg^2$]{Shen2013}. We choose to avoid box repetition, which would increase the light-cone sky area but result in correlated correlation function errors. The smaller area of the mock has an impact on the clustering measurements, particularly at small scales ($\la 0.5 \,h^{-1} \rm \, Mpc$), where the expected number of galaxy/QSO pairs is small \citep[see Table 2 of ][]{Shen2013}. The lack of clustering signal for ($r_p\la 0.5 \, h^{-1} \rm \, Mpc$) in Figure \ref{fig:wp_shen} may also indicate small-scale physics for the activation of SMBHs in galaxies that are missing from the current version of the mock. 

Central and satellite halos of the simulation are populated with CMASS galaxies using the HOD parametrisation of \cite{Shen2013}. We account for the CMASS selection function using the observed number-density of CMASS galaxies as a function of redshift, ${\rm d} N/{\rm d} z$ \citep[e.g.][]{Saito2016}. The mock CMASS galaxy sample is binned in redshift slices of ${\rm d} z =0.02$. We use a probabilistic approach to select mock CMASS galaxies so that their ${\rm d} N/{\rm d} z$ distribution matches the observed one. The CMASS galaxies are typicaly associated with massive halos, $\ga 10^{13}\,h^{-1} M_{\odot}$ \citep[e.g.][]{White2011}, and therefore the resolution of the MDPL2 simulation (particle mass $1.5\times10^9 \, h^{-1} \, M_{\odot}$) is sufficient to reproduce their clustering properties. 

Dark-matter halos in the light-cone are populated with AGN following the methodology described in Section \ref{sec:method}. The SDSS-DR7 QSO sample used by \cite{Shen2013} is essentially limited to the optical magnitude $i\approx19.1$\,mag for redshifts $z<1$. We account for this selection by assigning optical fluxes to individual sources as described in Appendix \ref{appendix:lc_vipers}. We select unobscured (type-I) AGN with optical magnitudes $15<i<19.1$\,mag. The bright limit is imposed by the SDSS QSO-target selection to avoid saturation and cross-talk in the spectra \citep{Schneider2010}.

\section{Redshift distributions of mock AGN}\label{appendix:dndz}

\begin{figure*}
\begin{center}
\includegraphics[height=1.\columnwidth]{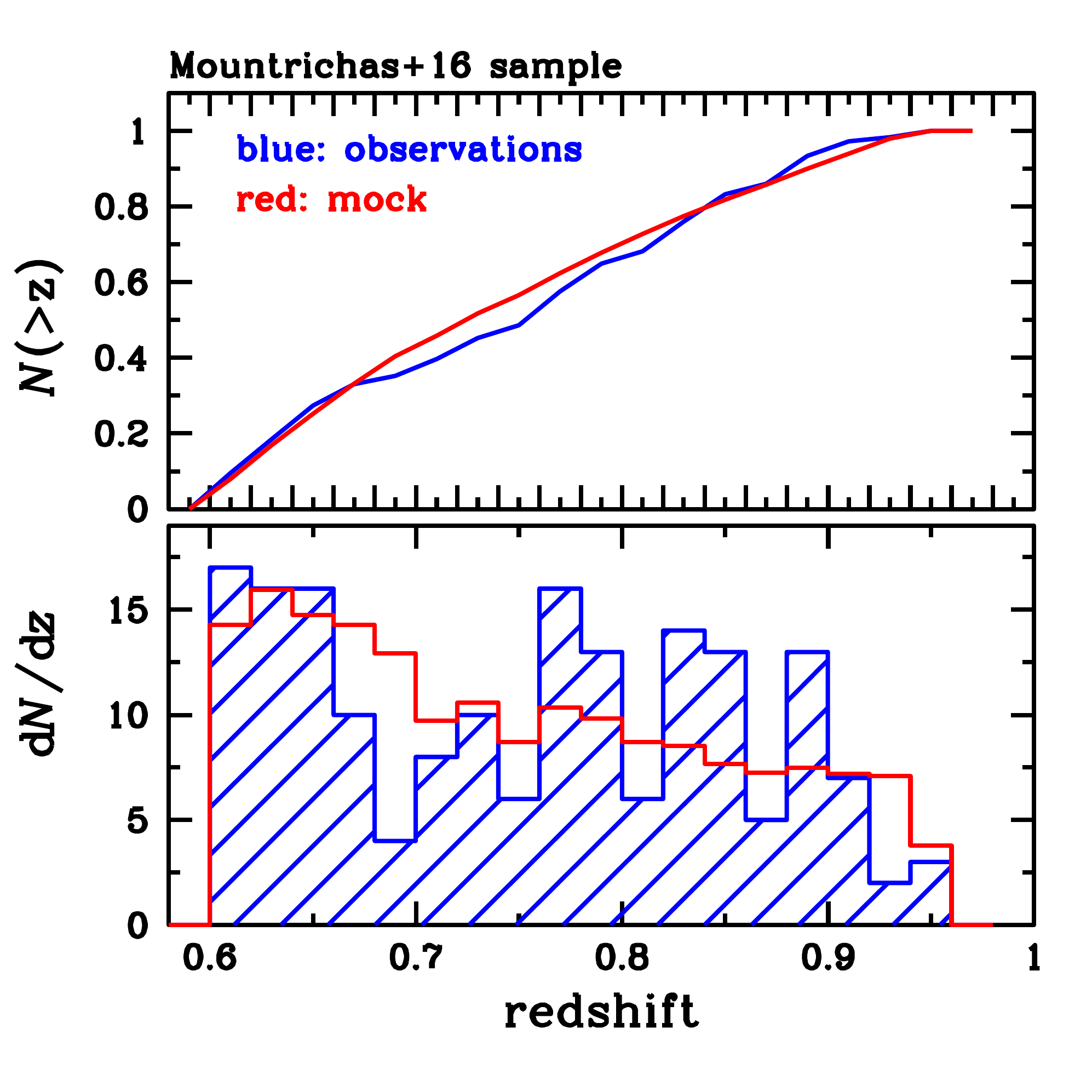}
\includegraphics[height=1.\columnwidth]{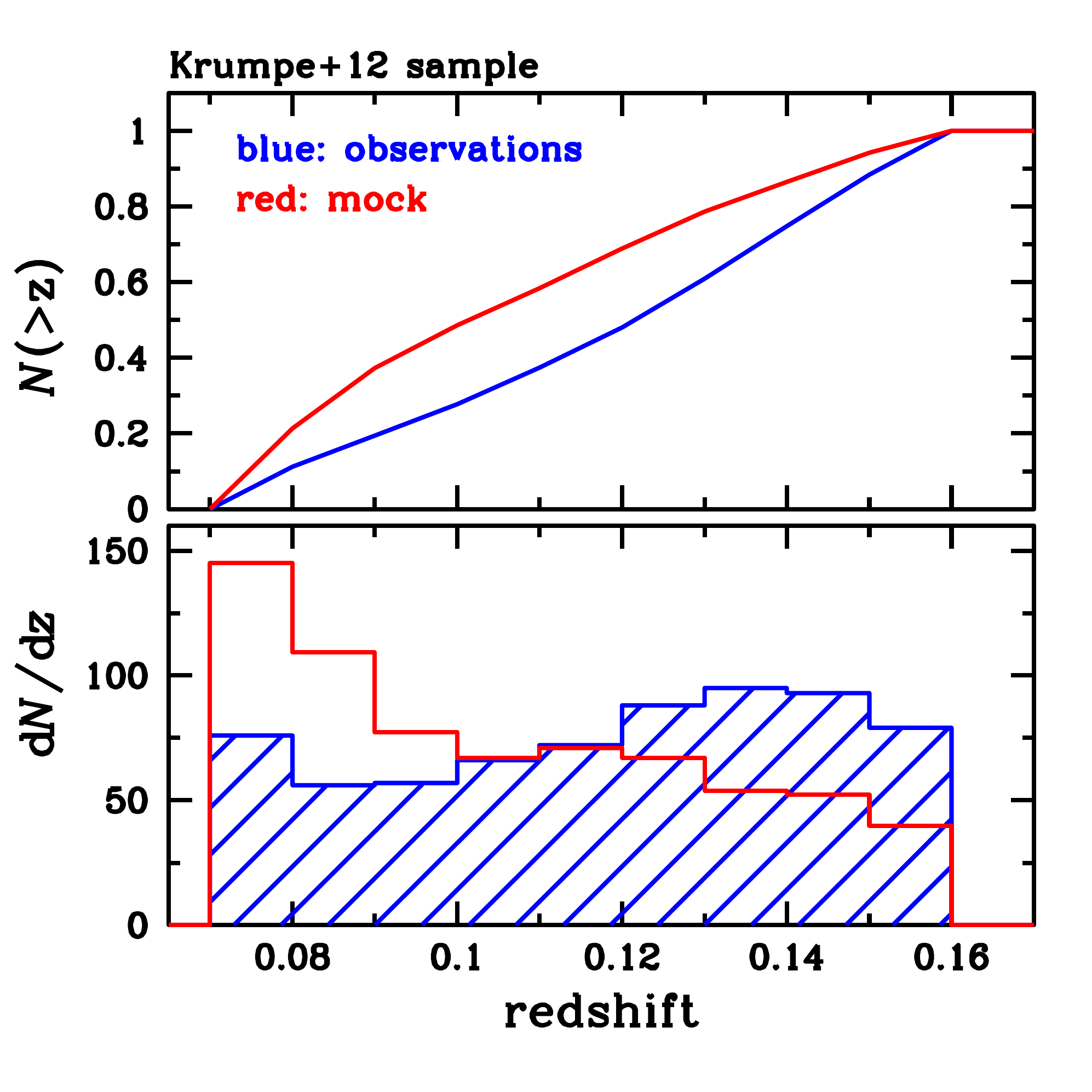}
\end{center}
\caption{Redshift distribution of the mock AGN (red curves and histograms) in comparison with the corresponding observed samples (blue curves and blue hatched histograms). The top set of panels plots the cumulative distributions, while the lower panels show the corresponding differential distributions. For the latter the observational and model histograms are normalised to the same area. The simulations corresponding to the \protect\cite{Mountrichas2016} sample (Appendix \ref{appendix:lc_vipers}) are shown on the left. These are compared with the redshift distribution of the AGN sample presented by \protect\cite{Mountrichas2016}  in the interval of the mock light-cone $z\approx0.6-1.0$. The KS-test shows that the null hypothesis that the the observed and mock distributions are drawn from the  same parent population cannot be rejected at a statistically significant level. The simulations corresponding to the \protect\citep{Krumpe2012} sample (Appendix \ref{appendix:lc_sdssmain}) are shown on the right. The mock in
  this case predicts a higher fraction  of AGN at the low-redshift end of    the    distribution     compared    to    the    observations.}\label{fig:dndzmock}
\end{figure*}

\begin{figure}
\begin{center}
\includegraphics[height=1.\columnwidth]{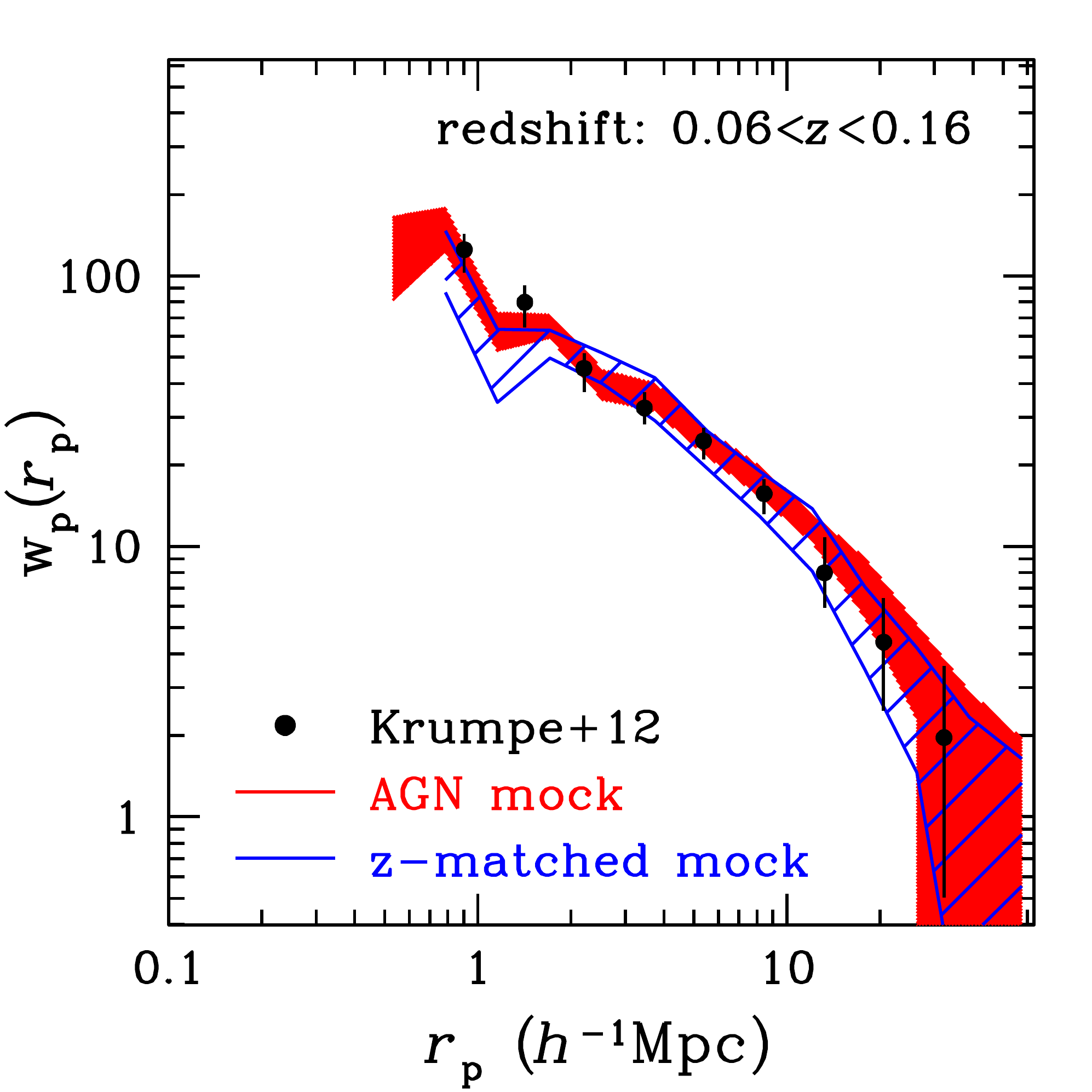}
\end{center}
\caption{The projected cross-correlation function of the RASS AGN and the SDSS Main galaxy sample in the redshift interval $0.07<z<0.16$. The data-points are the observational results of \protect\cite{Krumpe2012}. The red shaded region corresponds to the simulated data described in the Appendix \ref{appendix:lc_sdssmain} of the paper. This is the same as the red shaded region plotted in Figure \ref{fig:wp_rasssdss}. The blue hatched region is the correlation function for the mock AGN of Appendix \ref{appendix:lc_sdssmain} after resampling their redshifts to match the observed redshift distribution of the RASS AGN used by \protect\cite{Krumpe2012}.  The width of the shaded/hatched regions corresponds to the $1\,\sigma$ uncertainties determined using jackknife resampling.}\label{fig:wpDNDZ}
\end{figure}

We acknowledge that the redshift distribution of the mock AGN samples described in the previous Appendix sections do not always match the corresponding observed ones. This is demonstrated in Figure \ref{fig:dndzmock}, which compares the mock and observed redshift distribution (both differential and cumulative) for the  \cite{Mountrichas2016} and \cite{Krumpe2012} AGN samples. For the former sample the mock light-cones described in Section \ref{appendix:lc_vipers} produce AGN with redshift distribution consistent with the observed one. For the latter sample, however, there is a clear excess in the mocks at the low-end of the distribution compared to the real data. This is likely because the selection function of the ROSAT sample used by \cite{Krumpe2012}  is more complex than that adopted in the mocks, i.e. a simple X-ray flux cut. The Eddington bias, variations in the X-ray spectral shape among sources, the identification of ROSAT sources with Broad-Line spectroscopic counterparts \citep{Anderson2007} are all factors that complicate the sample selection function beyond a simple flux cut. It is noted that this is not an issue for the galaxy samples described above, since the HOD approach adopted for populating halos with galaxies produces (often by construction) redshift distributions that are consistent with the observations. 

Redshift distribution differences are a potential issue because they may affect the estimated 2-point correlation functions. Nevertheless the redshift range of the samples used in our work is relative narrow. Therefore, the precise distribution within these  narrow redshift intervals is a second-order effect. We demonstrate this point by resampling the mock differential redshift distribtion, $N(z)={\rm d} N/{\rm d} z$, to match the observed one. This is accomplished by assigning to each mock source, $i$, at redshift $z_i$ a weight, which is the ratio of the observed and mock ${\rm d} N/{\rm d} z$ distributions,  i.e. $w_i = N_{obs}(z_i)/N_{mock}(z_i)$.   For each mock source a random number is generated. The source is rejected from the sample if the random number is larger than the weight $w_i$ at the redshift of the source. This procedure essentially forces the redshift selection function of the mocks to match that of the real data. We then re-estimate the projected correlation function of the new sample and compare with the results presented in the paper. This is shown in Figure \ref{fig:wpDNDZ} in the case of the \cite{Krumpe2012} sample. This figure demonstrates that the precise details of the redshift distribution within the relatively narrow redshift range of the sample has a small impact on the results, at least within the error budget of the current set of simulations. We have repeated this exercise for other samples, for which there are differences between the observed and mock catalogue  redshift distributions \citep{Mountrichas2012, Shen2013} and confirmed the same result.

\bsp	
\label{lastpage}
\end{document}